\let\ifcom\iftrue
\let\ifprlstyle\iffalse

\ifprlstyle
\documentclass[reprint,aps,prl,twocolumn,superscriptaddress,longbibliography]{revtex4-2}
\else
\documentclass[aps,prx,twocolumn,superscriptaddress,longbibliography,10pt,tightenlines]{revtex4-2}
\fi

\usepackage{env}
\usepackage{lipsum}

\hypersetup{
    colorlinks=true,
    linkcolor=black,
    citecolor=navy,
    urlcolor=navy,
    breaklinks=true,    
}

\ifcom
\newcommand{\com}[2]{\sidenote{{\textbf{#1}: #2}}}
\newcommand{\sidenote}[1]{\textcolor{red}{#1}}
\else
\newcommand{\com}[2]{}
\newcommand{\sidenote}[1]{}
\fi

\renewcommand{\eqref}[1]{Eq.~(\ref{#1})}
\newcommand{\CNOT}{\text{CNOT}}

\begin{document}

\title{Quantum remeshing and efficient encoding for fracture mechanics}
 \author{Ulysse Rémond}
 %
 \affiliation{EDF R\&D, 7 Bd Gaspard Monge, 91120 Palaiseau, France}
 \affiliation{Sorbonne Université, Observatoire de Paris, Université PSL, CNRS, LUX, F-75005 Paris, France}
 \author{Pierre-Emmanuel Emeriau}
 \affiliation{Quandela, 7 rue Léonard de Vinci, 91300 Massy, France}
 \author{Liam Lysaght}
 \affiliation{Quandela, 7 rue Léonard de Vinci, 91300 Massy, France}

 \author{Jean Ruel}
 \affiliation{EDF R\&D, 7 Bd Gaspard Monge, 91120 Palaiseau, France}

 \author{Joseph Mikael}
 \affiliation{EDF R\&D, 7 Bd Gaspard Monge, 91120 Palaiseau, France}

 \author{Kyryl Kazymyrenko}
 \affiliation{EDF R\&D, 7 Bd Gaspard Monge, 91120 Palaiseau, France}

\date{\today}

\begin{abstract}
    We present a variational quantum algorithm for structural mechanical problems, specifically addressing crack opening simulations that traditionally require extensive computational resources.
    Our approach provides an alternative solution for a relevant 2D case by implementing a parametrized quantum circuit that stores nodal displacements as quantum amplitudes and efficiently extracts critical observables.
    The algorithm achieves optimal nodal displacements by minimizing the elastic energy obtained from finite element method. The energy is computed with only a polylogarithmic number of measurements.
    Extracting relevant scalar observables such as the stress intensity factor is then done efficiently on the converged solution.
    To validate the scalability of our approach, we develop a warm start strategy based on a remeshing technique that uses coarse solutions to circumvent barren plateaus in the optimization landscape of the more refined problems.
    Our method has been experimentally validated on Quandela's photonic quantum processor Ascella and comprehensive numerical simulations demonstrate its scalability across increasingly complex quantum systems.
\end{abstract}

\maketitle

\section{Introduction}

Partial differential equations (PDEs) 
provide the mathematical foundation for modeling a wide range of physical phenomena, from fluid dynamics 
and heat transfer %
to electromagnetic wave propagation  
and structural mechanics%
.

In solid mechanics, the stationary Navier–Cauchy equation governs the equilibrium state of an elastic continuum. It serves as a key building block for fracture mechanics \cite{griffith1921vi} and thermo-hydro-mechanical coupling models \cite{Coussy2004}. Its extensions capturing nonlinear hyperelasticity \cite{Ramberg1943} are being studied for their industrially relevant applications \cite{Fontana25}.

This governing equation belongs to the class of second-order elliptic PDEs, which can be reformulated as an energy minimization problem \cite{love2013}. The Finite Element Method (FEM) \cite{TJRHughes} is particularly effective for approximating the unique solution by optimizing nodal variables over a discretized spatial domain.

Modeling the fracture process requires the introduction of geometric singularities, such as local shape discontinuities, to represent the initial cracks.
This is an inherently multiscale procedure that, for the hydraulic dam example, ranges from the millimeter-scale concrete aggregate ($10^{-3} m$) to the infrastructure size ($10^2 m$).
Even if the fracture mechanics may involve the estimation of only a single scalar quantity as an output, simulating multi-crack propagation with high precision remains an extremely demanding task for state-of-the-art methods: current capacities may handle up to $10^{13}$ degrees of freedom in the resulting linear algebra problem \cite{Rüde}, which tend to be insufficient in the 3D applications.

These challenges highlight the limitations of classical FEM for fracture mechanics, particularly its steep cost when the initial crack distribution is uncertain.
Quantum algorithms offer a promising alternative for these simulations. Recent progress in quantum approaches to PDEs reflects this trend, targeting fundamental bottlenecks in classical numerical methods
\cite{harrowQuantumAlgorithmSolving2009,childsQuantumAlgorithmSystems2017a,caoQuantumAlgorithmCircuit2013a,berryHighorderQuantumAlgorithm2014,bravo-prietoVariationalQuantumLinear2023a,demirdjianVariationalQuantumSolutions2022,huQuantumCircuitsPartial2024,lindenQuantumVsClassical2020,SurveyHHL,lloydQuantumAlgorithmNonlinear2020,liuVariationalQuantumAlgorithm2021a,lloydQuantumAlgorithmNonlinear2020,Sato,kroviImprovedQuantumAlgorithms2023}.

The Harrow-Hassidim-Lloyd (HHL) algorithm \cite{harrowQuantumAlgorithmSolving2009} initially promised an exponential speedup for solving linear systems of equations. 
However, its practical application to PDEs faces significant constraint: it assumes an efficient quantum encoding of both the system matrix and right-hand side vector. The latter is rather challenging for geometries containing crack singularities, rendering the original HHL algorithm impractical for such applications. 
Tailored quantum algorithms for specific problems have also been proposed \cite{caoQuantumAlgorithmCircuit2013a}, but these approaches require large fault-tolerant quantum computers, which remain beyond current capabilities.

Consequently, some research has been shifted towards Variational Quantum Algorithms (VQA) solving PDEs on Noisy Intermediate-Scale Quantum (NISQ) computers \cite{demirdjianVariationalQuantumSolutions2022,liuVariationalQuantumAlgorithm2021a,Sato}. 
Most existing work, however, targets one-dimensional problems and suffers from optimization challenges such as barren plateaus \cite{McClean2018,Cerezo2021}, where gradients vanish exponentially with quantum system size.

Our work addresses these limitations along several critical directions: 
we extend the scope to realistic two-dimensional cases that better capture physical systems;
we propose an efficient quantum encoding tackling a fracture mechanics problem;
we experimentally validate the algorithm by running it on a photonic quantum device \cite{maring2024versatile};
and we develop a novel cascading warm-start strategy based on solutions from coarser meshes. Since the number of qubits directly corresponds to mesh refinement, our hierarchical approach enables solutions from rougher meshes to guide the initialization for finer meshes, mitigating the barren plateau problem and supporting higher-resolution simulations.

\section{Results}
\subsection{Mechanics of fracture in a pre-cracked plate}
\label{sec:res:fra}
\begin{figure}[H] 
    \includegraphics[width=0.45\textwidth]{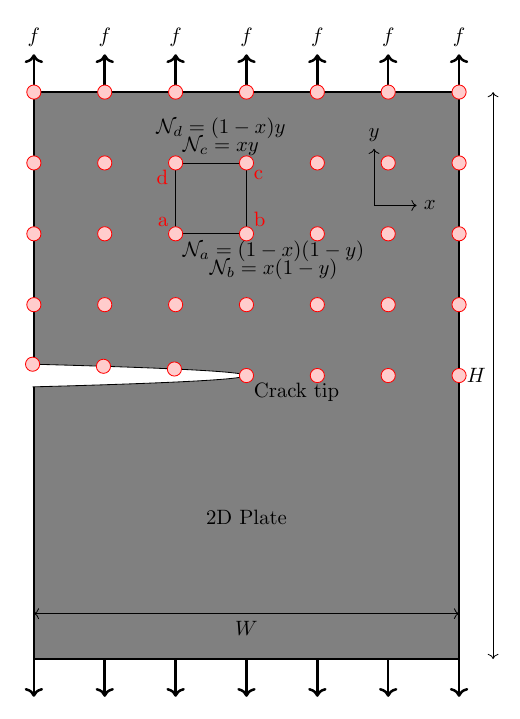}
    \caption{Geometry of the pre-cracked 2D plate under external loading, showing the crack tip and nodal grid. 
    This mesh underlies the FEM formulation (Eq. 3) and the operator decomposition (Eq. 5).}
    \label{fig:struct_fissuree}
\end{figure}

Linear elastic fracture mechanics studies the conditions under which a crack in a material will propagate. Here we focus on the standard example of a pre-cracked 2D plate (see Fig.\ref{fig:struct_fissuree}). System deformation evolves according to the static equilibrium linear Navier-Cauchy equation \cite{love2013}:
\beq\label{eq:Navier-Cauchy}
    \text{grad} (\text{div}\,\depl)+(1-2\nu)\Delta \depl= 0 
\eeq
where $\depl=(u_x,u_y)$ is the two components vector displacement field, $\nu\in[0,1/2)$ is the Poisson's ratio of the material, $\Delta$ is the Laplacian. 
Near the crack tip (Fig.\ref{fig:struct_fissuree}), the displacement gradient diverges \cite{irwin1957}; hence, the leading-order term dominates the solution and can be extracted. The corresponding scalar coefficient, called Stress Intensity Factor (SIF) \cite{westergaard39}, is therefore determined from the crack opening profile close to the tip:
\begin{equation} \label{eq:u2SIF}
    u_y(r) \approx \text{SIF}*2(1-\nu^2)\sqrt{2r/\pi},\quad \text{for }r\approx 0
\end{equation}
where $r$ is the distance from the crack tip.
In general, the SIF admits no closed-form expression for arbitrary geometries and boundary conditions. It is therefore tabulated for canonical cases or estimated numerically in specialized handbooks \cite{Murakami}.

For more complex cases, discretization schemes such as Finite Differences or the Finite Elements Method (FEM) are commonly used to solve \eqref{eq:Navier-Cauchy} 
 and to estimate numerically the SIF's value \cite{BrownSrawley}.
In FEM, the displacements are computed only on a grid of nodes and interpolated between them.
We use a regular mesh of $N_x\times N_y$ nodes of commensurate dimensions $N_x =2^{n_x}$ and $N_y =2^{n_y}$, so that the total number of degrees of freedom (DoFs) can be written as $N = 2\cdot N_x\cdot N_y\equiv 2^n$, where $n =n_x+n_y+1$ accounts for both components of the nodal displacement.
The mechanical deformation problem is converted into a finite-size optimization of the elastic energy $\mathcal{E}_{\text{\tiny fe}}(\ddepl)$, that is equivalent to the linear algebra problem \cite{TJRHughes}:
\begin{equation}
\label{eq:energy_FE}
    \min_{\ddepl}\mathcal{E}_{\text{\tiny fe}}(\ddepl) =  \min_{\ddepl} \left[\frac{1}{2}\ddepl \cdot \mathbb{K} \ddepl -\ddepl \cdot \vec f\right] \quad \Leftrightarrow \quad \mathbb{K} \ddepl = \vec f
\end{equation}
where $\ddepl \in \mathbb{R}^{N}$ is the vector of unknown nodal displacement corresponding to the discretized problem, $\vec f\in \mathbb{R}^{N}$ is the vector of external forces, and $\mathbb K\in\mathcal{M}_{N}(\mathbb R)$ is the stiffness matrix.

While $\vec{f}$ captures all variety of Neumann type boundaries, the Dirichlet conditions can be also enforced either through a complementary penalization contribution to the initial $\bbK$-matrix or by a projection technique that we cover in the next section.

The matrix $\bbK$, introduced in \eqref{eq:energy_FE}, is computed via FEM and is related to the gradient overlap of the displacement interpolation functions between adjacent nodes. 
Although $\mathbb{K}$ is sparse, its naive expansion into Pauli operators $\{X,Y,Z\}$ scales with the number of nonzero elements \ie effectively exponential in $n_x$ and $n_y$. This would render a direct quantum implementation inefficient.
Exploiting mesh symmetries, we obtain one of our main results: a polynomial-size decomposition of $\mathbb{K}$ into tensor products.
Indeed, if we define $\mathcal{S}$ as:
\begin{equation}
\label{eq:setS}
   \begin{array}{lll}
    X =
    \begin{pmatrix}
        0&1\\
        1&0
    \end{pmatrix}\quad  
    &Y=
    \begin{pmatrix}
        0&-i\\
        i&0
    \end{pmatrix} \quad 
    &Z=
    \begin{pmatrix}
        1&0\\
        0&-1
    \end{pmatrix}\\
    \mathcal{S}=\{p_\pm,\sigma_\pm, I\}\quad
    &p_\pm=
    (I\pm Z)/2   \quad 
    &\sigma_\pm=
    (X\pm iY)/2,
\end{array}
\end{equation}
the $\bbK$ matrix can be rearranged in a linear combination of tensor products $\{\bbM_i\}$ of matrices from $\mathcal S$.
\begin{equation}
    \label{eq:decomp:ph} \bbK = \sum_k \mu_k\bbM_k,\text{ with } \bbM_k = \bigotimes\limits_{l=0}^{n-1}m_{k,l}, \quad m_{k,l} \in \mathcal S,
\end{equation}
where, $\mu_k \in \mathbb R$ is a set of decomposition coefficients.
The derivation and explicit decomposition of $\mathbb K$ is detailed in Methods \ref{sec:met:fra:ten}.
Crucially, the number of terms in the sum in \eqref{eq:decomp:ph} scales logarithmically with the system size, that is as $O(n_x n_y)$.

We will now explore how the decomposition \eqref{eq:decomp:ph} paves the way to efficiently solving the discretized version \eqref{eq:energy_FE} of \eqref{eq:Navier-Cauchy} on a quantum computer.

\subsection{Solving efficiently elastic differential equations on a quantum computer}\label{sec:res:algo}

We now reformulate the minimization problem \eqref{eq:energy_FE} in quantum terms. A trial displacement vector is encoded as a parametrized quantum state $\ket \psi = \mathcal{A} \ket 0$, where $\mathcal{A}$ is the ansatz unitary. 
As usual with VQAs \cite{liuVariationalQuantumAlgorithm2021a, Sato}, the aim is to optimize the ansatz parameters so that it converges towards the target quantum state $\ket{\psi_{\text{tgt}}} = {\norm \ddepl}^{-1}\sum_i u_i \ket{i}$, where $u_i$ are the components of the solution $\ddepl\in\mathbb{R}^N$ of the discretized problem \eqref{eq:energy_FE} and $\ket{i}$ are the $2^n$ basis states of $n$-qubit quantum system, defined later in \eqref{eq:encoding}. 

We provide the quantum counterpart for the external force vector, that is supposed to be normalized: $\ket{f}=\sum_i f_i\ket{i}$, as explained later in \eqref{eq:f_circuit}.  Given the fact that the ansatz unitary generates a normalized amplitude vector in $\mathbb{C}^N$, the straightforward transcription of the cost function \eqref{eq:energy_FE} can be done only in the presence of at least one unrestricted rigid body motion, i.e. when $\det{\mathbb{K}}= 0$:
\begin{equation}
\label{eq:energy_q}
    \mathcal{E}_{\text{q}} (\psi)  = \frac{1}{2}\bra{\psi} \mathbb{K}\ket{\psi} - \operatorname{Re}\braket{\psi |f}.
\end{equation}
In this case the normalization constraint is absorbed into irrelevant kernel subspace movement. 
Otherwise, if $\det{\mathbb{K}}\neq 0$, the normalization coefficient should be explicitly introduced in the cost function. 
The norm optimization leads to a new form of normalization-agnostic quotient cost function \cite{Sato}:
\begin{equation}
\label{eq:energy_quot}
    \mathcal{E}_{\text{qq}} (\psi)  = -\frac{(\operatorname{Re}\braket{f | \psi})^2}{2\bra{\psi} \mathbb{K}\ket{\psi}}.
 \end{equation}
Regardless of whether \eqref{eq:energy_q} or \eqref{eq:energy_quot} is used, evaluating the cost requires only two quantities: $\bra{\psi} \bbK \ket\psi$ and $\braket{f | \psi}$.
We compute $\bra{\psi} \bbK \ket\psi$ by decomposing $\bbK$ into a polylog number of $\bbM_k$ matrices (see \eqref{eq:decomp:ph}). Indeed, only computations of the generic single qubit observable terms $\braket{\psi|\bbM_k|\psi}\equiv\braket{\psi|m_{k,0}\otimes m_{k,1}\otimes ...\otimes m_{k,n-1}|\psi}$ are needed, where $m_{k,l}\in \mathcal{S}$ and $k \in \{0,\dots,2(n_xn_y + n_x + n_y)\}$, as explained in Methods \ref{sec:met:fra:ten}.

We can estimate each term of this decomposition using common diagonalization techniques. 
While $p_\pm$ are already diagonal, the non-trivial observable measurements in the $\mathcal{S}$ basis are tensor powers of $\sigma_\pm$. Noticing that each such $\sigma_\pm$ term is present with its transposed counterpart, we focus on $\bra{\psi} \sigma_+^{\otimes n}+\sigma_-^{\otimes n} \ket\psi$  as an example.
We remark that the $\CNOT$-gate 
maps one non-hermitian operator $\sigma_+$ within a $\sigma_+^{\otimes n}$ term into a diagonal projector $p_+$.
More precisely, for two-qubit systems it is easy to verify that $\CNOT\cdot\sigma_+\otimes\sigma_+\cdot \CNOT= \sigma_+\otimes p_+ $, for three qubits $I\otimes \CNOT\cdot\sigma_+\otimes\sigma_+\otimes\sigma_+\cdot I\otimes \CNOT = \sigma_+\otimes \sigma_+\otimes p_+ $ and so on. 
These $(n-1)$ inductively-arranged $\CNOT$-gates can reduce $\sigma_+^{\otimes n}$ to $ \sigma_+\otimes p_+^{\otimes (n-1)}$, and similarly $\sigma_-^{\otimes n}$ to $\sigma_-\otimes p_+^{\otimes (n-1)}$ gates. This $\CNOT$-chain rotates $\ket{\psi}$ into $\ket{\tilde{\psi}}$,  leading to the following easily-diagonalizable operator estimator:
\begin{equation}
    \bra{\psi} \sigma_+^{\otimes n} +\sigma_-^{\otimes n} \ket\psi = \bra{\tilde{\psi}} \left(\sigma_++\sigma_-\right)\otimes p_+^{\otimes (n-1)} \ket{\tilde{\psi}},
\end{equation}
where $\ket{\tilde{\psi}}=\prod_{i=0}^{n-2} \CNOT_{i,i+1}\ket{\psi}$ is an inductively transformed quantum state; the $\CNOT_{i,i+1}$ operation applies the $\CNOT$ operator to two adjacent qubit. Noting that $X=\sigma_++\sigma_-$, the last step of diagonalization is done with the help of the Hadamard $H$-gate using the identity $Z=H\cdot X \cdot H$. Because $X\cdot\sigma_+\cdot X=\sigma_-$, all the others non-trivial tensor polynomials can be reduced to the previously discussed case.

Photonic computers implement the non unitary projection $p_\pm$, and thus also $\sigma_\pm = p_\pm X$, as elementary gates. Indeed in dual rail encoding, it amounts to herald the success of the projection on the relevant optical mode, that is measuring one mode and not detecting a photon there. 
As a consequence, $\bra{\psi} \bbK \ket\psi$ can alternatively be computed as a sum of photonic counts of the zero state $\bra 0$ after the circuit $\mathcal{A}^\dagger \bbK \mathcal{A} \ket 0$, which allows us to compute the estimator without the need for $\CNOT$-chains.

Furthermore, we provide an explicit quantum circuit for generating the boundary force vector $\ket{f}$. To do so, we define the ordering of the degrees of freedom such that:
\beq \label{eq:encoding}\ket{\psi_{\text{tgt}}}=\frac1{\norm \ddepl}\sum_{i=0}^{2^n-1}u_i\ket{i} = \sum_{x,y,d}u(x,y,d)\overset{n_y}{\overbrace{\ket y }\otimes}\overset{n_x}{\overbrace{\ket x}}\otimes\overset{1}{\overbrace{\ket d}},\eeq
where $u(x,y,d)$ is a shorthand notation for the normalized components of discretized displacement. Any grid index $0\le i_g <N_x\times N_y$ is represented in binary by a pair of coordinate indexes also written in binary $x,y$ ($i_g\equiv yx$). The additional bit $d$ separating horizontal $(d=0)$ and vertical $(d=1)$ displacement components is also introduced, which allows any qubit state to encode some given degree of freedom $\ket{i}=\ket{y,x,d}\equiv\ket{y}\otimes\ket{x}\otimes\ket{d}$.

An explicit representation of the unitary operator that generates an homogeneous Neumann boundary at the top of the plate (see Fig.\ref{fig:struct_fissuree}) is given by:
\begin{equation}
    \label{eq:f_circuit}\ket f = X^{\otimes n_y} \otimes H^{\otimes n_x}  \otimes X \ket0,
\end{equation}
allowing us to compute directly $\braket {f| \psi}$. %
For more complex boundary conditions, one may rely on the additive decomposition of the external boundaries or use a Walsh Series Loader \cite{Zyl23}.

\begin{figure*}[t]
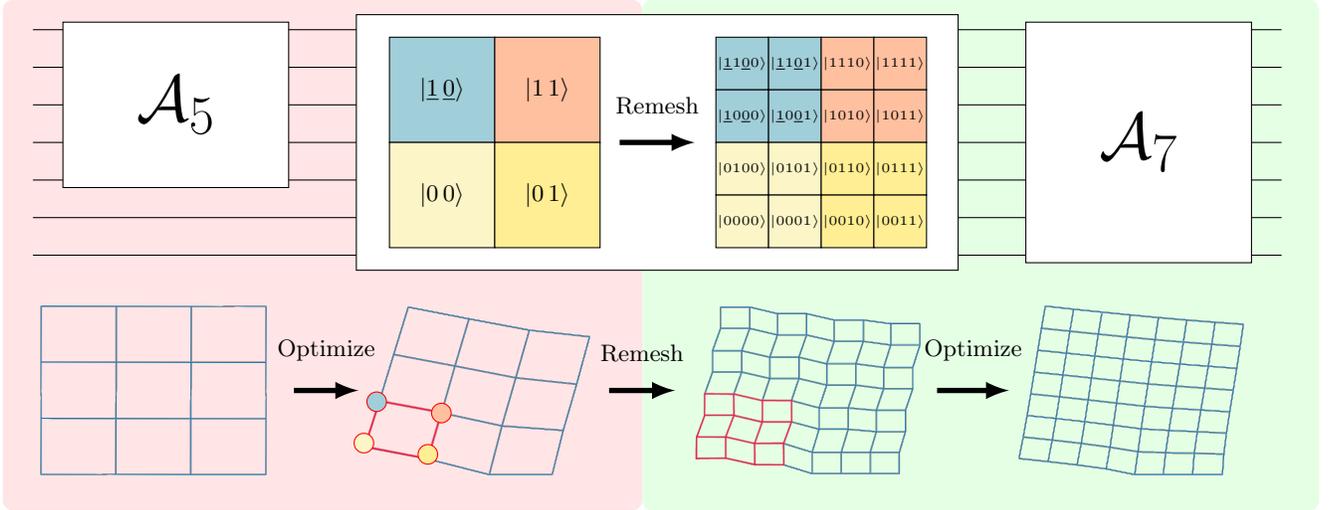

    \include{Figures/remeshing/5to7}
    \caption{We first optimize a 5-qubit state, and duplicate it to use it as a warm start for the 7-qubit ansatz. This scheme is applied iteratively until the desired system size is reached as seen in Fig. \ref{fig:c5719}.}
    \label{fig:nature57}
\end{figure*}

The finite element stiffness matrix $\bbK$ (given in \eqref{eq:K_decomp}) transcribes a purely Neumann boundary problem resulting in $\det \bbK=0$. Dirichlet boundaries greatly enrich the variety of treated situations, and eliminate spurious rigid body motions. We therefore consider the case of perfect clamping, where some DoFs are set to zero. 
It can be formally written with the help of the hermitian projector $\mathbb{P}\in \mathcal{M}_N$, as $\mathbb{P}\ket{\psi}=0$. We enforce it in the quantum system through a systematic projective restriction of all trial states, equivalent to the stiffness matrix modification: $\bbK\mapsto(I^{\otimes n}-\mathbb{P})\bbK(I^{\otimes n}-\mathbb{P}) + \mathbb P$.
The decomposition of any projector $\mathbb P$ within the $\mathcal{S}$ basis set is directly derived from the binary DoF's position encoding: $p_+\leftrightarrow 0$ and $p_-\leftrightarrow 1$, the last bit serving the choice between the $u_x$ and $u_y$ DoF's components.

\begin{equation}\label{eq:clamping_generic}
    \mathbb{P}= \overset{\text{y coord to block}}{\{p_{\pm}\}^{\otimes n_y}}\otimes \overset{\text{x coord to block}}{\{p_{\pm}\}^{\otimes n_x}}\otimes \overset{u_x\text{ or }u_y}{p_{\pm}}.
\end{equation}

The plate shown in the Fig. \ref{fig:struct_fissuree} exhibits a symmetry with respect to the crack plane. We model only the upper half, with the horizontal centerline constrained against vertical displacement. Moreover, we restrict horizontal rigid body movement by clamping the $u_x$ component of the crack tip.
It can be assessed via the following additive contributions for the full state projector $\mathbb{P}$:
\begin{align*}
p_{+}^{\otimes n_y}\otimes p_{-} \otimes I^{\otimes n_x -1}\otimes p_{-}  &\Leftarrow \text{ $u_y=0$ on the centerline,}\\
p_{+}^{\otimes n_y}\otimes p_{-} \otimes p_+^{\otimes n_x -1}\otimes p_+ 
&\Leftarrow\text{ $u_x=0$ on the crack tip.}
\end{align*}

Finally, we give circuit implementations for two common physically-relevant observables: SIF, defined in \eqref{eq:u2SIF}, and crack opening displacement (COD). With chosen numbering order \eqref{eq:encoding}, the COD-observable is estimated by multiplying the quantumly-computed norm \cite{Sato} with the amplitude of probability of zero counts in the optimized state $\ket{\psi}$:
\beq\text{COD}=\braket{0|\psi}\times \norm \ddepl \text{, where} \norm \ddepl = \frac{\operatorname{Re}\braket{f|\psi}}{\braket{\psi|\mathbb K|\psi}}\frac{1}{2^{n_x/2}} .\eeq

We similarly compute the SIF from the measurement of displacement amplitudes on the crack lips.
In the vicinity of the crack tip the solution for the crack opening scales as a square root of the distance from the crack tip \eqref{eq:u2SIF}.
It could allow one to calculate the SIF from a single displacement value. However, 
the single-point measure would be highly error-prone. On the one hand, close to the crack tip the displacement amplitude value is vanishing necessitating an important shots number to eliminate statistical noise, on the other hand, far from the crack tip the asymptotic solution deviates from its theoretical expression \eqref{eq:u2SIF}. Therefore, we opt for a multiple point estimator and give here an integral measurement that averages the SIF-value over the whole crack lip of length $W/2$, see Fig.\ref{fig:struct_fissuree}:
\begin{equation}\label{eq:SIFQ}
   \text{SIF} = \frac{\sqrt{\pi} \norm \ddepl}{(1-\nu^2)W}\sqrt{\int_0^{W/2}  \left[u(x,y=0,d=1)\right]^2dx}
\end{equation}
The discretized value of the integral can once again be computed with the help of the projector $\mathbb{P}=p_{+}^{\otimes n_y}\otimes p_{+} \otimes I^{\otimes n_x -1}\otimes p_{-}$:
\begin{align}\label{eq:SIFQ_projector}
    &\int_0^{W/2}  \left[u(x,y=0,d=1)\right]^2\,dx \approx \frac{W}{2^{n_x}} \braket{\psi|\mathbb{P}|\psi}, \nonumber \\ 
    &\text{SIF} = \frac{\sqrt{\pi\braket{\psi|\mathbb{P}|\psi}} \cdot\operatorname{Re}\braket{f|\psi}}{(1-\nu^2)\sqrt{W}\braket{\psi|\mathbb{K}|\psi}}\frac{1}{2^{n_x}}
\end{align}

Alternatively, for large-size systems, a more precise measurement can be obtained by restricting the integration domain closer to the crack tip. For $x\in [\frac{3W}{8},\frac W2]$, the projector $p_{+}^{\otimes n_y}\otimes p_{+}\otimes p_{-}^{\otimes 2} \otimes I^{\otimes n_x -3}\otimes p_{-}$ should be used. 

Interestingly, during the optimization both physical observables (COD and SIF) converge to their targets much faster than the cost function itself, due to the high energy density concentration close to the crack tip ---see Fig. \ref{fig:obs_casca}.

\subsection{Warm start with coarse mesh solution}

\indent Barren plateaus pose a major challenge in VQA optimization. They represent regions where the cost function's gradient vanishes across most parameter variations.
Having defined a global \cite{Cerezo2021} cost function, we observe a flat landscape for the random state initialization in sufficiently large systems ($n\gtrsim10$), hampering optimization. %
In order to avoid this difficulty, we derive a quantum version of the mesh refinement technique \cite{Ern2004}.
This approach paves the way towards consistent warm start creation, i.e. problem-agnostic strategies to bypass barren plateaus in PDEs.

The idea is to define two discretized  versions (coarse and refined) of the same continuous problem and to use the converged coarse $n$-qubits solution as the initial state for the refined $(n+2)$-qubits solution. This strategy can be repeated iteratively until we reach the desired precision, hence a \textit{cascaded} approach. %
Since barren plateaus are not an issue for a low number of qubits, we can always compute the target function from a cold start (i.e. using a randomly chosen initial state) for the coarsest meshes.%

Once we have solved the problem for a mesh, we duplicate the mesh, increasing the number of qubits --- one additional qubit for each $x$ and $y$ coordinates to maintain the elements' aspect ratio. The basis state written in binary $\ket{y}\ket{x}$ is extended to $\ket{y}\ket{x}\ket{0}\ket{0}$. %
Then, all the data are duplicated, since each newly added qubit is transformed into $\ket + = H\ket 0$ with a Hadamard gate.
Finally, we enforce the order of qubits (see Sec. \ref{sec:met:sim:sca}
) creating new extended coordinates along both directions, %
so that the initial coarse state $\ket{y}\ket{x}$ is mapped to the refined one $\ket{y}\ket +\ket {x}\ket +$. %
In summary, the procedure can be illustrated as:\\ $\ket{y}\ket{x}\overset{+2}{\mapsto}\ket{y}\ket{x}\ket{00}\overset{H}{\mapsto}\ket{y}\ket{x}\ket{+}\ket{+}\overset{swap}{\mapsto}\ket{y}\ket{+}\ket{x}\ket{+}.$   %

Any smooth function represented by a $n$-qubit state turns into the piecewise-constant function represented by this $(n+2)$-qubit state, see Fig. \ref{fig:curvedup1D}. This procedure is similar to a homogeneous remeshing with constant field extrapolation on the nearest neighbors. %
The interpolation of the original displacement field introduces local field discontinuities, shifting the $(n+2)$-cost function from the target value already achieved at the previous step of $n$-qubit optimization. This error induced by the field projection is well known and has been studied in depth in the classical remeshing technique \cite{Ern2004}. %

Third, we further fine-tune this displacement field with another $(n+2)$ parametrized circuit until convergence is achieved, resulting in the local field relaxation commonly called the equilibration phase after field projection.

\begin{figure}[htbp]

    \begin{minipage}[b]{0.47\textwidth}
    \subfloat[\centering Cascaded warm-start remeshing strategy.\label{fig:c5719}
    ]
    {%
    \includegraphics[width=\textwidth]{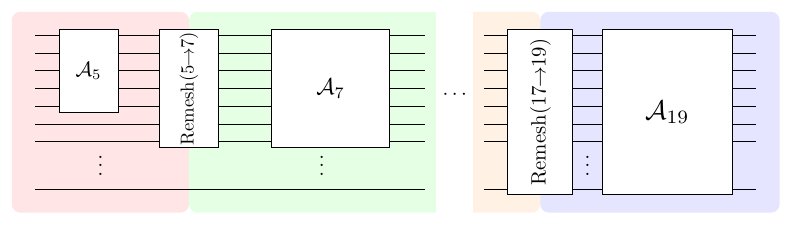}
    }\\[1em]
    \subfloat[\centering Evolution of observables with increasing discretization. Convergence ratio of remeshing-based and cold-start results with regard to the target continuous values.\label{fig:obs_casca}]
    {%
      \includesvg[width=\textwidth]{Figures/obs_19_cold.svg}%
    }\\[1em]
    \subfloat[\centering 2H one-step optimization, comparing remeshing-based and cold-start results with the same-size classical solutions, showing the average, best, 1st and 9th deciles.  \label{fig:obs_dev_2H_19Q}]
    {%
      \includesvg[width=\linewidth]{Figures/obs_dev_2H_ENE_2e-3.svg}%
    }
  \end{minipage}
  \centering
  \caption{{Cascaded simulations from 5 to 19 qubit mesh. In (b) the iterative remeshing, performing 48h optimization at each step, allows to reach 19 qubit simulation with $50\%$ precision of observable COD and SIF, far beyond the capacities of any tested cold start strategies. 
(c) illustrates a single step remeshing capabilities to bypass the initial barren plateau with 50 Layers ansatz. The $n+2$-qubit state is optimized for 2 hours starting from an ideal state before duplication, obtained from $n$-qubit discretization.
}}
\end{figure}

This whole three-step process can be repeated (Fig. \ref{fig:c5719}), increasing the mesh refinement up to the desired precision. In doing so, we obtain successive warm starts to avoid barren plateaus at each step as shown in Fig. \ref{fig:obs_casca}.%

We test numerically this quantum remeshing procedure.
We achieved quantifiably better results than the cold start strategies. Initialized from 5 qubits onwards, systems composed of up to $5 \cdot10^{5}$ DOFs, i.e. 19 qubits, were successfully simulated with this method using the circuit in Fig. \ref{fig:c5719}. 

As shown in Fig. \ref{fig:obs_casca}, even for low energy convergences at the most refined simulations, the cascading remeshing procedure proved especially reliable in reaching satisfactory convergence for the relevant mechanical observables: SIF (60\%) and COD (70\%). Cold start simulations had error ratios of more than 100\% from 13 qubits onwards. The SIF and COD precision peaks near 9 qubits as the coarser value suffer from classical discretization errors, and the finer values suffer from a limited VQA convergence time and the lack of an tailor-made ansatz.

Fidelity stays high (over 87\%) throughout the whole cascading procedure, and optimizations using it as cost function yield unrealistic physical deformations. This illustrates how even quantum states with high fidelity might not correspond to a physically-acceptable state, justifying the need of more relevant quantities such as the elastic energy to be used as cost function.

The respective convergences of such observables in 2 hours from the classical solution at the former step can be assessed in Fig. \ref{fig:obs_dev_2H_19Q}.

\FloatBarrier
\subsection{Experimental implementation}

We experimentally validated a 4-qubit version of our variational quantum algorithm on Quandela’s linear optical QPU \cite{maring2024versatile}. Our ansatz uses the QLOQ compression scheme \cite{lysaght2024quantum} which enables efficient encoding of multi-qubit states onto photonic hardware. The parametrized quantum circuit was encoded on two photons, with each run consisting of $10^5$ shots.
Ascella is a photonic quantum processor that utilizes a semiconductor quantum dot based on-demand single photon source \cite{somaschi2016near}, with two-photon interference visibility above 94\%. Up to 6 single photons can be fed synchronously into every odd input of a 12-mode universal interferometer implemented in a photonic integrated circuit. The interferometer is fully reconfigurable, enabling the implementation of $12\times12$ unitary transformations with high fidelity -- 99.7\% amplitude fidelity for Haar random 12x12 unitaries \cite{fyrillas2024high}. A threshold single photon detector system measures the photon distribution and coincidences at the output of the optical circuit.

\begin{figure}[h!tb]
    \centering
    \includegraphics[width=0.45\textwidth]{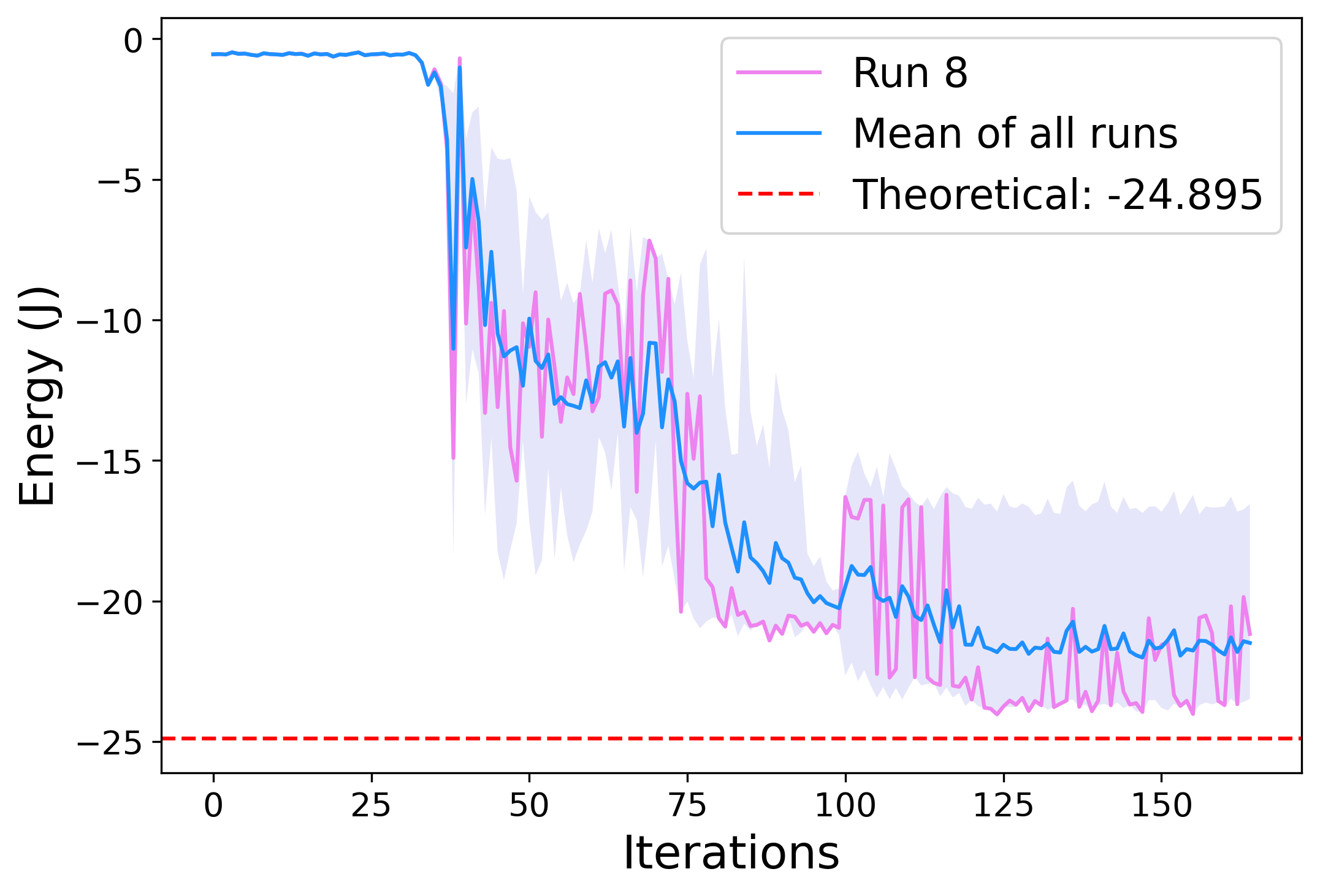}
    \caption{The results from 10 runs on Quandela’s linear optical QPU. The blue line shows the mean energy estimate at each iteration, while the shaded region is bounded by the minimum and maximum values. Run 8, which produced the minimum energy estimate overall, is shown as a pink line.} 
    \label{fig:QPU_plot}
\end{figure}

10 QPU runs are displayed in Figure \ref{fig:QPU_plot} to show that this is working on average. The mean final energy estimate of these QPU runs was $-23.15$, with a best result of $-24.060 \pm 0.003$ while the target value is $-24.895$. To further enhance accuracy, basic error mitigation techniques were applied during the runs (see Sec. \ref{sec:met:fra} for details). This performance aligns with the best-case scenario predicted by noiseless simulations, which reached $−23.952$.

\FloatBarrier
\section{Discussion}

Our results demonstrate the feasibility of addressing computationally intensive structural mechanics problems—specifically fracture mechanics—using VQAs. 
By formulating the 2D Navier-Cauchy equation as an elastic energy minimization problem, we circumvent the impractical requirements of algorithms like HHL while providing a clear route to scalability with the developed warm start strategy. 
To our knowledge, this work constitutes the first implementation of a two-dimensional elasticity problem with realistic boundaries on a quantum platform, marking an important step beyond prior studies, which were limited to one-dimensional PDEs, simpler boundaries or scalar cases.
The proposed encoding, which stores nodal displacements as quantum amplitudes, allows efficient computation of critical observables such as the stress intensity factor (SIF) and crack opening displacement (COD) through simple measurements.

A major contribution of this work is the development of a quantum remeshing technique that systematically mitigates barren plateaus --one of the primary obstacles in VQA optimization. By initializing finer discretizations from optimized coarser meshes, we demonstrated a significant improvement in convergence over cold-start and simple warm-start strategies. We numerically validated this hierarchical warm-start approach up to 19 qubits, illustrating scalability to physically meaningful regimes. Moreover, we find that physical observables converge much faster than the energy functional, suggesting that even partially optimized states can yield accurate engineering quantities of interest.

The experimental validation on a photonic quantum processor confirms the practical implementability of our approach on real hardware. Despite noise and device imperfections, our results show we achieve energy estimates within $\approx96\%$ of the theoretical optimum with basic error mitigation, closely matching noiseless simulations.

Looking ahead, several avenues merit exploration. First, the generalization to three-dimensional domains and vector-valued PDEs beyond elasticity (e.g., thermal and hydraulic diffusion, linear themo-mechanical coupling...) is straightforward within our encoding scheme and would significantly broaden the applicability of this method. Second, further improvements of the optimization procedure, possibly through problem-inspired ansätze, tensor-network-inspired decompositions and tailored classical minimization algorithms, could enable simulations on larger systems and with more complex geometries. Finally, given the modular nature of the algorithm, it offers a robust benchmarking tool for quantum hardware.

Overall, this work establishes a concrete pathway for leveraging quantum computing in high-fidelity structural simulations. 
By coupling hierarchical VQA strategies with scalable encodings and efficient observable estimation, and by demonstrating the first 2D case on real hardware, we move one step closer to practical quantum advantage in computational mechanics.

\section{Methods}
\subsection{Fracture Mechanics}
\label{sec:met:fra}

The present paper examines the second order Navier-Cauchy differential vector-equation \cite{Lame1833} that naturally arises in the isotropic linear elasticity theory under hypothesis of small transformations. In the 2D case under additional plane strain hypothesis, the mechanical equilibrium of the structure may be written as a set of two constraints on the two components vector displacement field  $\depl = (u_x,u_y)$, where $u_\alpha=u_\alpha(x,y)$ for $\alpha=\{x,y\}$ is a scalar displacement in the direction of one of its euclidean coordinates $\alpha$ at the given position $(x,y)$ in space. The Navier-Cauchy equilibrium problem in the absence of volumetric density forces (i.e. weight) then reads as a constant coefficients linear system for the two components displacement field $\depl = (u_x,u_y)$:
\begin{align}\label{eq:Navier-Cauchy_component}
    \begin{cases}
    \partial_x(\partial_x u_x +\partial_y u_y) + (1-2\nu)(\partial^2_x +\partial^2_y) u_x=0, \\
    \partial_y(\partial_x u_x +\partial_y u_y) + (1-2\nu)(\partial^2_x +\partial^2_y) u_y=0;
    \end{cases}
\end{align}
where $\nu$ is the Poisson's ratio of material. Commonly used boundary conditions in mechanics often involve a combination of Dirichlet-type conditions, such as perfect clamping ($\depl(x,y)=0$ for $ (x,y) \in \partial \Omega_D$) and  Neumann-type constrains on the gradient of displacement $\partial \Omega_N = \partial  \Omega \setminus \partial \Omega_D$. 

The problem can be rewritten as a variational minimization of the elastic density functional $\mathcal{E}$. In the case of full Dirichlet condition constraining $\depl(x,y)$ to be equal to $\depl_D(x,y)$ on the entire boundary $\partial \Omega$, $\mathcal{E}$
is a quadratic form: %
\begin{align}
\label{eq:meca_vari_2D}
    \mathcal{E} [\depl]  = \frac{1}{4}\int_{\Omega} &(\partial_x u_x+\partial_y u_y)^2 \text{d}\Omega\, + \\ + \frac{1-2\nu}{4}\int_{\Omega} & \left[(\partial_x u_x-\partial_y u_y)^2 +(\partial_x u_y+\partial_y u_x)^2\right] \, \text{d}\Omega. \nonumber
\end{align}
Any displacement field minimizing \eqref{eq:meca_vari_2D} and satisfying this Dirichlet boundary respects the Euler-Ostrogradski extrema conditions 
that are equivalent to the initial set of equations \eqref{eq:Navier-Cauchy_component}.

In the presence of Neumann conditions, a linear term is added to \eqref{eq:meca_vari_2D}.
This modification is presented in \eqref{eq:energy_FEfull}, albeit in the discretized case.

\eqref{eq:Navier-Cauchy_component} being parametrized by a single coefficient $\nu$, 
the solution's complexity stems from either the boundary conditions or the sample geometry.

Theoretical solutions for the sample with a sharp-edged geometry exhibit a divergent gradient of displacement in the vicinity of this geometric singularity \cite{westergaard39}. The leading order in this divergent asymptotic expansion is called the stress intensity factor (SIF),  \eqref{eq:u2SIF}. SIF quantifies the threshold beyond which crack propagation occurs, characterizing the critical structure's loading. It is thus widely used in Linear Elastic Fracture Mechanics theory (LEFM) \cite{griffith1921,irwin1957}.
Another typical singularity-describing observable is the crack opening displacement (CO)\cite{Murakami}, which reveals the deformation level. While the displacement field $\depl$ is a primary variable of interest for general mechanical problems, fracture mechanics mainly focuses on the computation of aforementioned pair of observables on each singularity. 
The CO is a local pointwise measurement, whereas the SIF is obtained via integration over a domain close to a singularity.

In the current work, we use a quantum variational algorithm to numerically solve the classical 2D LEFM problem of a pre-cracked plate (Fig. \ref{fig:struct_fissuree}), 
subjected to a uniform density of force $f$
on its upper and lower bounds, 
neglecting the off-plane movement. 

\subsection{Numerical discretization}

As in any linear PDE, the Navier-Cauchy equation admits an approximate discretized solution. In the finite difference method the differential operators are estimated on a grid (see Fig.\eqref{fig:struct_fissuree}), resulting in a linear algebraic equation for the vector of nodal unknowns $\ddepl$: $\mathbb{K}\ddepl= \vec f$. If $N_g$ is the total number of grid nodes (whereas $N=2N_g$ is the number of DoF), $\ddepl = (u_{ix},u_{iy})_{i=1}^{N_g} \in \mathbb{R}^N$ is a discretized approximation of the continuous displacement field $\depl$, i.e. the finite dimensional vector of nodal displacement in both horizontal ($x$) and vertical ($y$) directions. The stiffness matrix $\mathbb{K}$ is large, but usually sparse.

In the more subtle finite element method (FEM), the continuous minimization problem is converted into a finite size optimization as the trial displacement field is reduced to a linear combination of element by element piecewise continuous polynomials: $\depl(x,y)=\sum_{i=1}^{N_g} \mathcal{N}_i(x,y) \depl_i $. As each basis function $\mathcal{N}_i$ is normalized to unity at its reference node $i$ vanishing outside the adjacent elements region, all the expansion coefficients represent the nodal values of displacement field: $\depl_i\equiv\depl(x_i,y_i)$. The minimization of the discretized energy \eqref{eq:meca_vari_2D} becomes a finite dimensional algebraic problem, see also \eqref{eq:energy_FE}:%
\begin{align}
\label{eq:energy_FEfull}
    \mathcal{E}_{\text{\tiny fe}} (\ddepl)  = \frac{1}{2}\sum_{i,j=1}^{N_g} \sum_{\alpha,\beta}u_{i\alpha} u_{j\beta} \mathbb{K}_{i\alpha j\beta} 
    - \sum_{i=1}^{N_g}\sum_{\alpha} u_{i\alpha}f_{i\alpha} 
\end{align}
Any Neumann-type boundary is represented by some vector of external forces $\vec f\in \mathbb{R}^{N}$; and the Greek letters $\alpha,\beta$ run over Euclidean coordinates $\{x,y\}$. All coefficients $\mathbb{K}_{i\alpha j \beta}$ of a FEM's stiffness matrix are related to a generic gradient correlator expression: $\int_{\Omega} \partial_\alpha \mathcal{N}_i \partial_\beta \mathcal{N}_j \, \text{d}\Omega$.

The minimization of discretized energy potential $\mathcal{E}_{\text{\tiny fe}}(\ddepl)$ by variation of the nodal displacements amplitudes is once again                                                                  equivalent to resolving the algebraic linear equation.

\subsection{Tensor product decomposition of the stiffness matrix}
\label{sec:met:fra:ten}

In order to estimate the $N^2$ coefficients of $\mathbb{K}$ matrix, all cross-integrals of the gradient of shape functions $\int_{\Omega} \partial_\alpha \mathcal{N}_i \partial_\beta \mathcal{N}_j \, \text{d}\Omega$ should be calculated on the whole $\Omega$, scaling as $\sim N^2$. For large $N$, the FE software rely on a more efficient assembling algorithm, that scales linearly is system size.  First, the domain is split into a set of simple polygonial subdomains, called elements: $\Omega= \bigcup_{el}\Omega_{el}$. Second, for each element, all the connected stiffness contributions are computed, giving for quad4 mesh a $4\times4$ elementary matrix.
In what follows, we extend this procedure with the aim of decomposing the stiffness matrix into a more primitive form of tensor product of $2\times 2$ matrices more suitable for quantum calculations.

The whole domain integrals can be separated into the sum of elementary contributions:
\begin{align}\label{eq:dNdN_element}
    \int_{\Omega} \partial_\alpha \mathcal{N}_i \partial_\beta \mathcal{N}_j \, \text{d}\Omega = \sum_{el}\int_{\Omega_{el}} \partial_\alpha \mathcal{N}_i \partial_\beta \mathcal{N}_j \, \text{d}\Omega; 
\end{align}
We consider here a rectangular $N_x\times N_y$ domain, composed of linear Quad4 quadrangles \cite{dhatt2012,r3.01.01}. Here $N_x$, or $N_y$ is the number of discretization nodes in $x$ (or $y$ direction). Since for this regular mesh all the elements have the same geometry, all the elementary integrals in $\mathbb{K}$ are identical. It allows us to separate the nodal $(i,j)$ connectivity from the $(x,y)$ component's links stored in $2\times2$ matrices $\bbK_{ij}$. These elementary contributions are only dependent on their local connectivity index pairs. With local index notations $i,j\in\{a,b,c,d\}$ (see Fig.\ref{fig:struct_fissuree}) each matrix term is given by the integral over the Quad4 quadrangle:

\begin{widetext}
\begin{equation}\label{eq:Kij_all}
    \mathbb{K}_{ij}\equiv \int_\text{quad}\text{d}\Omega
    \begin{pmatrix}
    (1-\nu) \partial_x \mathcal{N}_i \partial_x \mathcal{N}_j  + (1-2\nu)/2 \partial_y \mathcal{N}_i\partial_y \mathcal{N}_j&
    (1-2\nu)/2\partial_y \mathcal{N}_i\partial_x \mathcal{N}_j  +  \nu \partial_x \mathcal{N}_i \partial_y \mathcal{N}_j \\ 
    (1-2\nu)/2\partial_x \mathcal{N}_i\partial_y \mathcal{N}_j  +  \nu \partial_y \mathcal{N}_i \partial_x \mathcal{N}_j  & 
    (1-\nu) \partial_y \mathcal{N}_i \partial_y \mathcal{N}_j  + (1-2\nu)/2 \partial_x \mathcal{N}_i\partial_x \mathcal{N}_j
  \end{pmatrix}  
\end{equation}
\end{widetext}

\begin{figure}
    \centering
    \includegraphics[width=\linewidth]{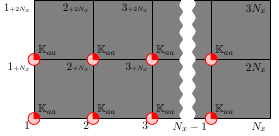}	
	\caption{$\mathbb{K}_{aa}$ component contribution.}
	\label{fig:factorK_aa}
\end{figure}

All expressions for $\bbK_{ij}$ can be found in Appendix \ref{sec:appendix:Kij}. Let us first analyse the contribution of an on-site $\mathbb{K}_{aa}$ term. 
An initially intuitive idea is to set it as a purely diagonal block matrix. However, the expected diagonal nature of the on-site term comes from the sum of all the four $\sum_{i\in\{a,b,c,d\}}\mathbb{K}_{ii}$ contributions. For boundary nodes, the summation is only partial. The bottom-left corner node, number $1$ on (Fig.\ref{fig:factorK_aa}), has exclusively the $\mathbb{K}_{aa}$ on-site contribution. As the total number of elements in the mesh is equal to $(N_x-1)(N_y-1)$, $\mathbb{K}_{aa}$ appears then in the global rigidity matrix the same number of times, namely once during the integration of each quad4 element. The whole set of top and the most right-hand side nodes is not coupled through the $\mathbb{K}_{aa}$ term. This fact is reflected in the block diagonal structure of the full rigidity matrix $\mathbb{K}_{aa}$ contribution, that reads with an appropriate row-by-row nodal numbering as:
\begin{equation}\label{eq:K_aa_full}
\mathbb{D}_{N_y}\otimes\mathbb{D}_{N_x}\otimes\mathbb{K}_{aa};    
\end{equation}
The $\mathbb{D}_{N_\alpha}$ for $\alpha\in \{x,y\}$  is the identity matrix with a missing last diagonal element -- see \eqref{eq:DTU}.

\begin{figure}
    \centering
    \includegraphics[width=\linewidth]{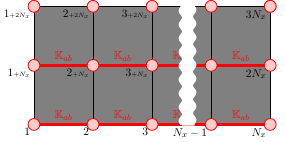}	
	\caption{$\mathbb{K}_{ab}$ component contribution.}
	\label{fig:factorK_ab}
\end{figure}
We analyse further the $\mathbb{K}_{ab}$ connectivity terms. For each row, the lower nodes are coupled between the nearest neighbors $N_x-1$ times. The upper row stays uncoupled. The contribution of the $\mathbb{K}_{ab}$ term to the full rigidity matrix of rectangular $N_x\times N_y$ shape sample therefore gives: $\mathbb{D}_{N_y}\otimes\mathbb{T}_{N_x}\otimes\mathbb{K}_{ab}$, where $\mathbb{T}_{N_x}$ is a upper single diagonal Toeplitz matrix -- see \eqref{eq:DTU}.

The calculation of each elementary contribution is similar to our previous development. By analogy with $\mathbb{D}_{N}$ we can introduce $\mathbb{U}_{N}$, the identity matrix without its first diagonal term, such that the full rigidity matrix aggregates into:
\begin{align}\label{eq:K_decomp}
    \mathbb{K}& 
    =\mathbb{D}_{N_y}\otimes\mathbb{D}_{N_x}\otimes\mathbb{K}_{aa} +
    \mathbb{D}_{N_y}\otimes\mathbb{U}_{N_x}\otimes\mathbb{K}_{bb} + 
    \nonumber \\&  
    \mathbb{U}_{N_y}\otimes\mathbb{U}_{N_x}\otimes\mathbb{K}_{cc} +
    \mathbb{U}_{N_y}\otimes\mathbb{D}_{N_x}\otimes\mathbb{K}_{dd} +
    \nonumber \\&
    \mathbb{D}_{N_y}\otimes\mathbb{T}_{N_x}\otimes\mathbb{K}_{ab} + 
    \mathbb{D}_{N_y}\otimes\mathbb{T}_{N_x}^\dagger\otimes\mathbb{K}_{ba} + 
    \nonumber \\& 
    \mathbb{T}_{N_y}\otimes\mathbb{D}_{N_x}\otimes\mathbb{K}_{ad} +
    \mathbb{T}_{N_y}^\dagger\otimes\mathbb{D}_{N_x}\otimes\mathbb{K}_{da} + 
    \\&
    \mathbb{T}_{N_y}\otimes\mathbb{U}_{N_x}\otimes\mathbb{K}_{bc} + 
    \mathbb{T}_{N_y}^\dagger\otimes\mathbb{U}_{N_x}\otimes\mathbb{K}_{cb} +
    \nonumber \\ &
    \mathbb{U}_{N_y}\otimes\mathbb{T}_{N_x}\otimes\mathbb{K}_{dc} +
    \mathbb{U}_{N_y}\otimes\mathbb{T}_{N_x}^\dagger\otimes\mathbb{K}_{cd} +
    \nonumber \\&
    \mathbb{T}_{N_y}\otimes\mathbb{T}_{N_x}\otimes\mathbb{K}_{ac} + 
    \mathbb{T}_{N_y}^\dagger\otimes\mathbb{T}_{N_x}^\dagger\otimes\mathbb{K}_{ca} + 
    \nonumber \\&
    \mathbb{T}_{N_y}\otimes\mathbb{T}_{N_x}^\dagger\otimes\mathbb{K}_{bd} + 
    \mathbb{T}_{N_y}^\dagger\otimes\mathbb{T}_{N_x}\otimes\mathbb{K}_{db}
    \nonumber ,
\end{align}
where all these matrices can be explicitly defined with index notation, as $\mathbb{D}_{N_\alpha}^{ij}=\delta_{ij}-\delta_{iN_\alpha}\delta_{jN_\alpha}$;
$\mathbb{T}_{N_\alpha}^{ij}=\delta_{i,j-1}$ and 
$\mathbb{U}_{N_\alpha}^{ij}=\delta_{ij}-\delta_{i1}\delta_{j1}\text{, for } i,j\in[1,N_\alpha]$.
\begin{equation}\label{eq:DTU}
    \begin{array}{ccc}
         
    \mathbb{D}_{N_\alpha} \in \mathcal{M}_{N_\alpha} &
    \mathbb{T}_{N_\alpha} \in \mathcal{M}_{N_\alpha} &
    \mathbb{U}_{N_\alpha} \in \mathcal{M}_{N_\alpha}\\
    
    \begin{pmatrix}
    1      & \hdots & 0      & 0      \\
    \vdots & \ddots & \vdots & \vdots \\
    0      & \hdots & 1      & 0      \\
    0      & \hdots & 0      & 0
    \end{pmatrix}\llap;&
    \begin{pmatrix}
    0      & 1      & \hdots & 0      \\
    \vdots & \ddots & \ddots & \vdots \\
    0      & 0      & \ddots & 1      \\
    0      & 0      & \hdots & 0
    \end{pmatrix}\llap;&
    \begin{pmatrix}
    0          &0      & \hdots & 0      \\
    0        &1      & \hdots & 0      \\
    \vdots   &\vdots & \ddots & \vdots \\
    0        &0      & \hdots & 1           
    \end{pmatrix}\
    \end{array}
\end{equation}

As the number of nodes in each direction is set to a power of two, i.e. $N_x=2^{n_x}$ and $N_y=2^{n_y}$, we can express these matrices using the $2\times2$ matrices of the spanning family $\mathcal S$ defined in \eqref{eq:setS}\cite{liuVariationalQuantumAlgorithm2021a}:
\beq
\begin{aligned}
    \mathbb{D}_{N_\alpha} = I^{\otimes n_\alpha} - p_{-}^{\otimes n_\alpha}; \qquad
    \mathbb{U}_{N_\alpha} = I^{\otimes n_\alpha} - p_{+}^{\otimes n_\alpha};\\
    \mathbb{T}_{N_\alpha} \equiv \mathbb{T}(n_\alpha)=\sum\limits_{k=0}^{n_\alpha-1} I ^{\otimes k} \otimes \sigma_+ \otimes \sigma_-^{\otimes n_\alpha-k-1}.
\end{aligned}
\eeq
Crucially, the number of quantumly-computable terms in this decomposition is logarithmically small and can be accessed via $2n_x n_y+ O(n_x)+O(n_y)$ independent single-qubit quantum measurements. In fact, each leading-order connection term ($\mathbb{K}_{ac}$ and $\mathbb{K}_{bd}$) counts for $4 n_x n_y$ $\mathcal{S}$-tensor product contributions, but their factorization yields directly 
diagonalizable matrices:
\begin{align*}
     &\left(\mathbb{T}(n_y) + \mathbb{T}^\dagger(n_y)\right)\otimes \left(\mathbb{T}(n_x) + \mathbb{T}^\dagger(n_x) \right)\otimes I, \\
     &\left(\mathbb{T}(n_y) - \mathbb{T}^\dagger(n_y)\right)\otimes \left(\mathbb{T}(n_x) - \mathbb{T}^\dagger(n_x) \right)\otimes X.
\end{align*}
Given that the matrices $\mathbb{T}(n_\alpha) \pm \mathbb{T}^\dagger(n_\alpha)$ are constructed from $n_\alpha$ single-qubit observables, the resulting leading-order scaling is $ 2n_x n_y$ as expected.

This scaling property reflects not only the sparse nature of the rigidity matrix itself, which has $\sim 2^n$ non zero elements, but also the fact that we use an overdetermined set of $2\times 2$ matrices for decomposition $\mathcal{S}$. 
The simultaneous usage of redundant $I\equiv p_++p_-$ matrix is absolutely crucial, as we remark that the simplest identity matrix expansion $I^{\otimes n}=(p_++p_-)^{\otimes n}$ has $2^n$ non-zero spanning components in $p_\pm$ basis.

\subsection{Application using Photonic Computing}
\label{sec:met:exp}
\begin{figure}[!ht]
    \centering
    \includegraphics[width=0.45\textwidth]{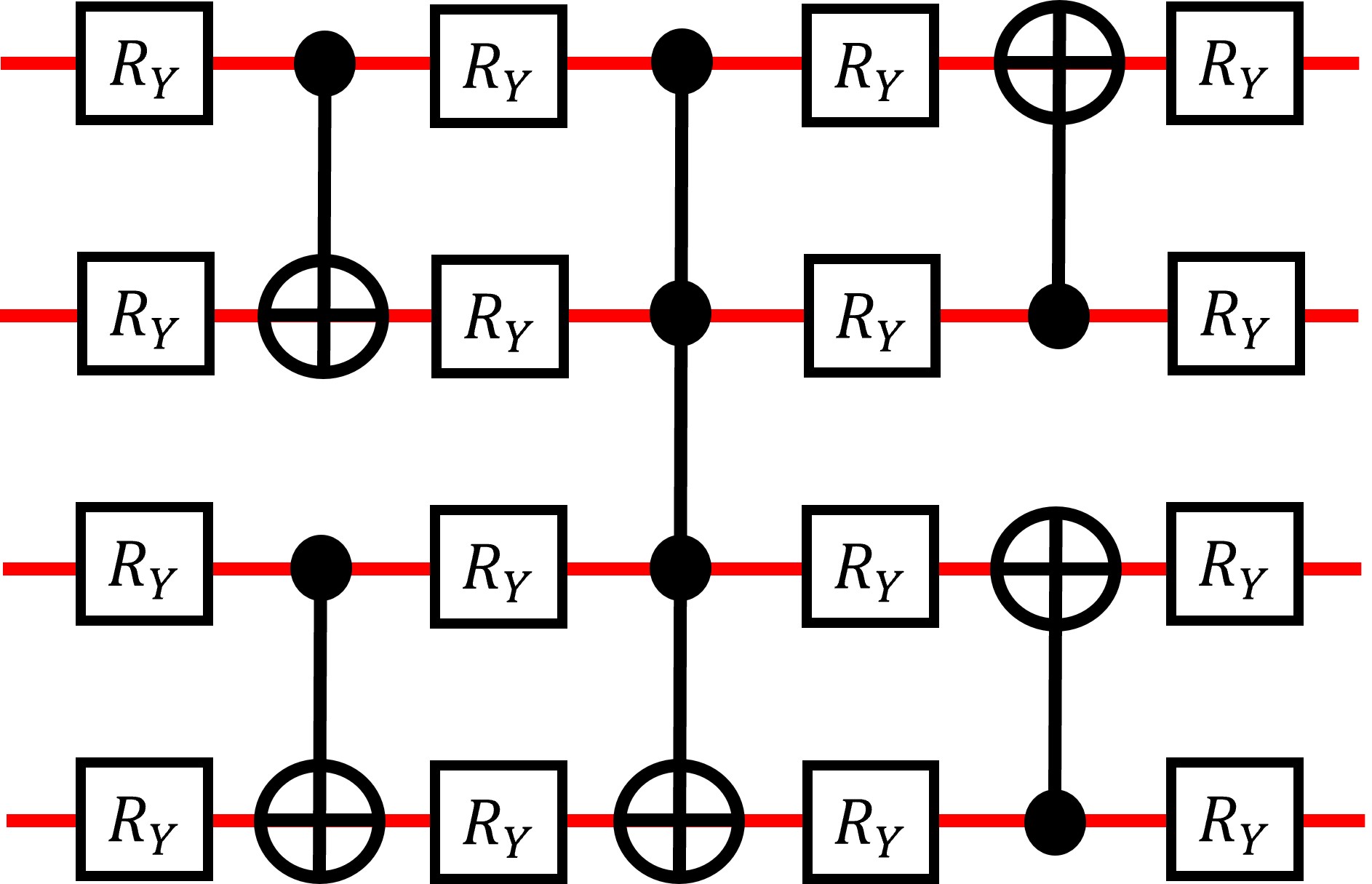}
    \caption{The parametrized quantum circuit used in our QPU runs and simulations. An unbalanced version of the CCCX gate was used, where the success probability varies depending on the input state (see Section F.1 of Lysaght et al 2024 \cite{lysaght2024quantum} for more details).
} 
    \label{fig:QPU_circ}
\end{figure}

The parametrized QLOQ circuit we implemented is displayed on Figure \ref{fig:QPU_circ}. It splits 4 qubits in two groups of 2 qubits that are encoded on a four-level system (4 modes with 1 photon) \cite{lysaght2024quantum}. 

To compensate for small device errors on the QPU we applied a noise mitigation technique from iteration 100 onwards in each run, whereby any value in the circuit’s outcome probability distribution below a certain threshold (0.001 in this case) was set to 0, and then the distribution was renormalised. This improved final energy values from approx -21 to approx -23. Applying this noise mitigation technique for all iterations prevents the ansatz from converging below approx -9, which is why it is only activated on iteration 100. 

To optimize the variational quantum circuit, we use the classical gradient-free optimiser NEWUOA \cite{powellNEWUOASoftwareUnconstrained2006} with the following stopping criteria: maximum 220 iterations, $5 \times 10^{-14}$ absolute tolerance on the result and $5 \times 10^{-10}$ relative tolerance on the result.

We note that the ansatz is not expressive enough to converge.
To verify this, we ran 1000 noise-free infinite-shot VQA simulations \cite{heurtel2023strong} with the same initial parameters. The mean final energy estimate of these simulations was $-23.77 \pm 0.3$, with a best result of $-23.95$. We chose this ansatz because it is the largest we can fit on the current optical chip \cite{lysaght2024quantum}.

The mean final energy estimate of the VQA was better in these noise free simulations than on QPU (Sim:-23.77 vs QPU:-23.15). However, the best QPU result was slightly better than the best noise free simulation result (Sim:-23.95 vs QPU:-24.06). This is likely due to the stochastic nature of variational algorithms on noisy hardware. 

Both the noisy simulations and QPU runs compare favourably to the theoretical value, with best simulated and QPU results achieving 96.2\% and 96.6\% accuracy respectively.

\subsection{Quantum mesh refinement}
\label{sec:met:sim:sca}

\begin{figure}[H]
    \centering
    \begin{adjustbox}{max size={\linewidth}{\textheight},center}
      \includegraphics{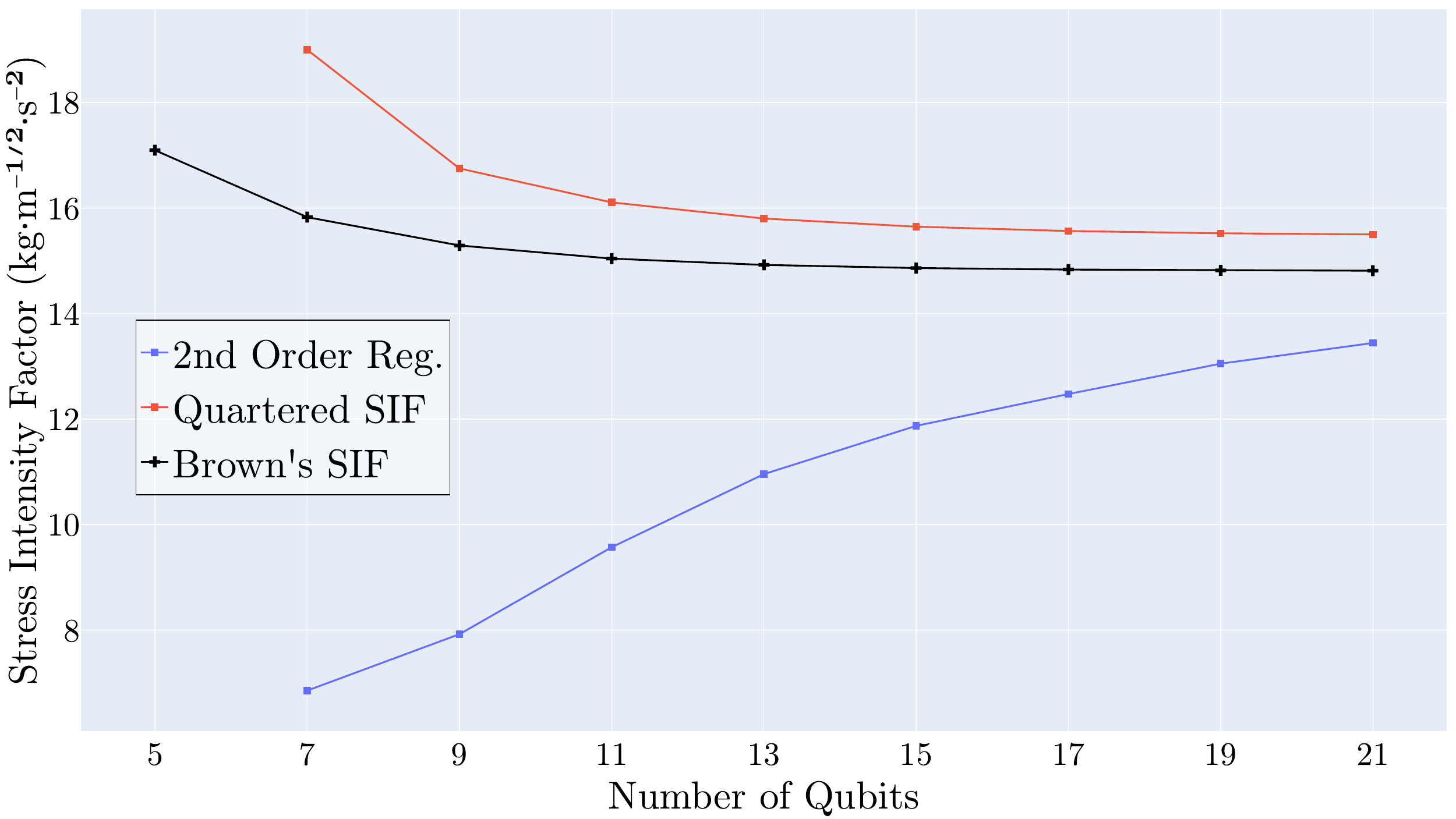}
    \end{adjustbox}
    \caption{Convergence of different ways to compute the SIF observable in $2D$-plate. The integration over crack lips \eqref{eq:SIFQ_projector} (red) and the polynomial interpolation (blue) are to be compared with the reference solution (black) from \cite{BrownSrawley}.}
    \label{fig:SIF:obs}
\end{figure}
\subsubsection{From classical to quantum remeshing}
Mesh refinement is a crucial technique in the FEM for improving the accuracy of numerical solutions. The elements where a higher resolution is needed are subdivided, typically in regions where the solution exhibits discontinuities, steep gradients, or stress concentration. The refinement process increases the number of DoFs, allowing the finite element solution to better capture fine-scale features. The coarse solution is usually projected on the refined mesh in order to get the new discretized field initialization.

In case of fracture mechanics, the precise estimation of the SIF necessitates either high refinement close to the crack tip or the usage of specific FE \cite{barsoum1976}. In the classical resolution, with the FEM, the uniform refinement is inappropriate, as the resulting increase in the number of DoF rapidly overcomes the classical computers capacities. Fig. \ref{fig:SIF:obs} illustrates this rather slow convergence of the SIF observable in the regular mesh hypothesis: the mesh is reaching $\sim 1$ million of DoFs (21 qubits) before a few percent error to the reference solution is attained even in the simple case of the pre-cracked plate. However, this classical system size bottleneck is alleviated through the exponential capacity of quantum memory. However, variational convergence in the resulting high-dimensional parameter spaces can be difficult.

\begin{figure}[htbp]
    \centering
    \includegraphics[width=\linewidth]{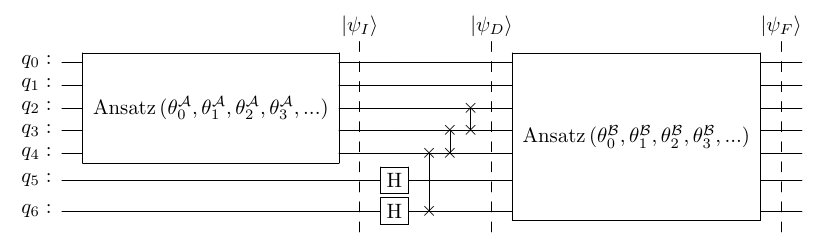}
    \caption{$6$-qubit circuit to be optimized by varying the $\theta^{\calB}$, starting from a $4$-qubit circuit that was already optimized by varying the $\theta^{\calA}$.}
    \label{fig:ansatz2DVEC_remeshed}
\end{figure}

Therefore, we propose to modify the standard VQA procedure incorporating an additional \textbf{quantum remeshing} step. The quantum solution $\ket{\psi_I}$ obtained for some given discretization level is first projected on the uniformly refined mesh: $\ket{\psi_I}\mapsto\ket{\psi_D}$. 
The coarse solution serves as the initial state for convergence in the refined space%
: $\calA\ket{\psi_D}\mapsto \ket{\psi_F}$ instead of the standard 'cold start' $\calA\ket{0}\mapsto \ket{\psi_F}$ (see Fig. \ref{fig:ansatz2DVEC_remeshed}). %
In what follows, we illustrate our proposition with some relevant examples.

\subsubsection{1D Remeshing.}\label{sec:cas:1D}

We first consider the $1D$ case, with a scalar field of unknowns $\depl(x)\in\mathbb{R}$. %
The quantum $n$-qubit state encodes $2^n$ DoFs $u(x)$ of the discretized field by the amplitudes of the ket $\ket x$, \ie $\ket{\psi_I}=\sum\limits_{x=0}^{2^n-1} u(x)\ket x$, where $x$ is written in binary as usual. $\ket{\psi_D}$ should represent the initial guess or the warm start for a more refined solution of $n+1$ qubits. 
As $\depl(x)$ is supposed to represent some continuous physical solution, we can obtain a good enough approximation of the remeshed function by duplicating the each $u(x)$ value on its right close nearest neighbor as shown in Fig \ref{fig:curvedup1D}:

\begin{figure}[H] 
    \centering
    \includesvg[width=0.49\textwidth]{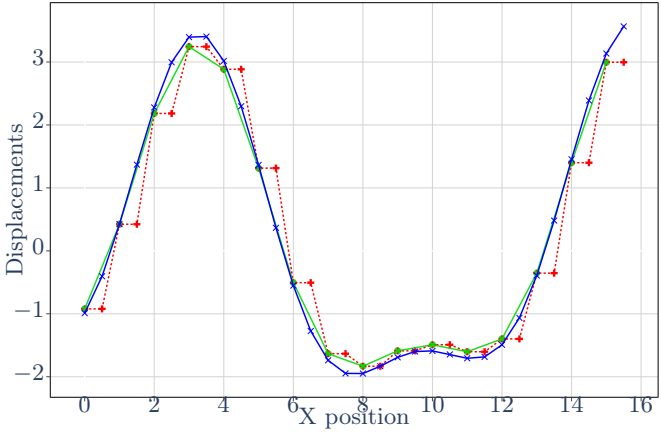}
    \caption{The 4-qubit solution $\ket{\psi_I}$ (green), the 5-qubit target state $\ket{\psi_F}$ (blue), and the 5-qubit state $\ket{\psi_D}$ obtained by duplication of that 4-qubit state (red)}
    \label{fig:curvedup1D}
\end{figure}

\begin{figure}[H]
    \centering
    \includegraphics[width=0.7\linewidth]{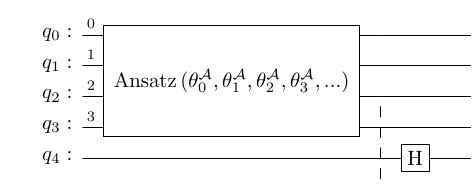}
    \caption{Circuit yielding the $\ket{\psi_D}$ state containing the duplicated displacements.}
    \label{fig:ansatz1D+wire}
\end{figure}

This duplicated state contains values for twice as many points as $\ket{\psi_I}$, but values for successive points are equal, that allows us to write it in the compact form using the Euclidean division ($//$): 
\begin{equation}
    \ket{\psi_D} = \sum\limits_{x=0}^{2^{n+1}-1} u(x // 2)\ket x. 
\end{equation}
Let us call $\calA$ the ansatz that yields the $n-$th qubit solution $\ket{\psi_I}$. To create the duplicated state, we then simply need to enrich this circuit by an extra little endian qubit and to apply the Hadamard gate in order to duplicate the data, see Fig. \eqref{fig:ansatz1D+wire}.

\begin{figure}[htbp]
    \centering
    \includegraphics[width=\linewidth]{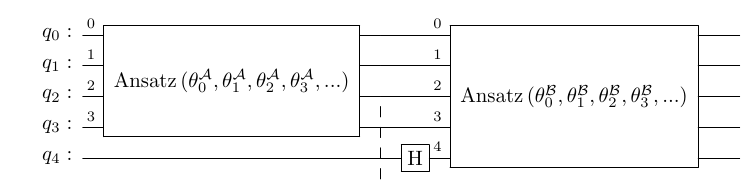}
    \caption{Circuit $\calC$ to be optimized stage by stage towards the target at $5$-qubit state}
    \label{fig:ansatz1D_remeshed}
\end{figure}

This causes the state to evolve as follows:
\beq\label{eq:addingh1D}
\begin{aligned}
\ket{\psi_D}&=(\iden^{\otimes n} \otimes H) \cdot \left( \ket{\psi}\otimes \ket 0 \right)
=\ket{\psi}\otimes \ket +\\
&= \sum\limits_{x=0}^{2^{n}-1} u(x)\ket {x}\otimes\left(\ket0 + \ket1\right)\\
&= \sum\limits_{x=0}^{2^{n}-1} u(x)\ket {2x} + \sum\limits_{x=0}^{2^{n}-1} u(x)\ket {2x+1}\\
&= \sum\limits_{x=0}^{2^{n+1}-1} u(x // 2)\ket x
\end{aligned}\eeq
We will now no longer optimize over the parameters $\theta^{\calA}_{i}$ of $\calA$ to limit the size of the parameter space.
To improve the state, another stage of $n+1$ qubit optimization is added via an ansatz $\calB$ as shown in Fig. \ref{fig:ansatz1D_remeshed}. We construct $\calB$ ansatz so that it realizes an identical transformation for all zero set of its parameters: $\calB(\theta=0)\equiv I^{\otimes(n+1)}$. By randomly setting the initial parameters of $\calB$ close to zero
, we optimize them from a so-called warm start, which greatly improves the convergence. We remark that this requested ansatz property can easily be enforced for any kind of ansatz by the additional application of the dagger-circuit: $\calB(\theta)\cdot\calB^\dagger(\theta=0)$.%

You can then repeat the procedure by considering this entire circuit $\calC$ represented in Fig. \ref{fig:ansatz1D_remeshed} like a black box, as we considered $\calA$ throughout this section. Each $q$-qubit state becomes the warm start for its $q+1$-qubit remeshed counterpart.

\subsubsection{2D Remeshing}\label{sec:cas:2D}
In $2D$, the construction is similar but we need to reorder qubits instead of simply appending them. %

\paragraph{Scalar fields}\label{sec:2D:sca}
For a scalar field 
we don't need any qubit to account for local DoF, 
like in Sec. \ref{sec:cas:1D}. We can therefore identify any DOF by its position $(x,y)$: quantumly speaking, a trial state $\ket{\psi_I}$ is proportional to $\sum\limits_{y=0}^{2^{n_y}-1}\sum\limits_{x=0}^{2^{n_x}-1} {u}(x,y)\ket {y,x}$, %
where $x$ and $y$ are written in binary. Notice how points are ordered first along the $y$ axis, and then along $x$, so that the next point of a point $(x,y)$, is $(x+1,y)$ if $x<N_x-1$, and $(0,y+1)$ otherwise. 
If we want to apply a remeshing along both directions $x$ and $y$, we must first obtain the following ``duplicated'' state:%
\beq\ket{\psi_D} = \sum\limits_{y=0}^{2^{n_y+1}-1} \sum\limits_{x=0}^{2^{n_x+1}-1} u(x//2,y//2)\ket {y,x}\eeq
To do so, not only do we have to use Hadamard gates to account for how the newest (little-endian) qubit should distribute its amplitude equally between $\ket0$ and $\ket1$ (as in Sec. \ref{sec:cas:1D}), but also add swap gates so that the new qubits reach the correct position in the numbering. This compels us to also add a reordering circuit as follows:%

\begin{figure}[htbp] 
    \centering
    \includegraphics[width=\linewidth]{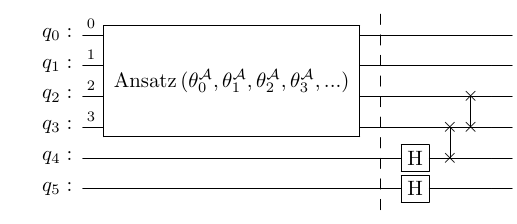}
    \caption{Circuit yielding the state $\ket{\psi_{2,D}}$ containing the duplicated displacements.}
    \label{fig:ansatz2D+wire}
\end{figure}
This causes the state to evolve as follows:
\begin{align}
\ket{\psi_2}\otimes \ket {00} &\to \ket{\psi_2}\otimes \ket {++} \tag*{after Hadamard\qquad}\\
&=  \sum\limits_{y=0}^{2^{q_y}-1} \sum\limits_{x=0}^{2^{q_x}-1} \ddepl_{\calA,q}(x,y)\ket {y,x} \otimes\ket {++} \nonumber\\
&= \sum\limits_{y=0}^{2^{q_y}-1} \sum\limits_{x=0}^{2^{q_x}-1} \ddepl_{\calA,q}(x,y)\ket{y} \ket {x++} \\
&\to \sum\limits_{y=0}^{2^{q_y}-1} \sum\limits_{x=0}^{2^{q_x}-1} \ddepl_{\calA,q}(x,y)\ket{y+}\ket {x+}\tag*{after swaps\qquad} \\
&= \sum\limits_{y=0}^{2^{q_y+1}-1} \ket{y} \sum\limits_{x=0}^{2^{q_x+1}-1} \ddepl_{\calA,q}(x//2,y//2)\ket {x} \quad \tag*{like in Eq. \ref{eq:addingh1D}\qquad} \end{align}

Finally, like in the 1D case, we simply have to compose this circuit with an ansatz that implements the identity for known parameters, and optimize starting from such parameters.%

\paragraph{Vector fields}\label{sec:2D:vec}

For a vector field with two local DoF, we need one additional qubit compared to the previous sections, as shown in \eqref{eq:encoding}.

An ansatz-created state corresponding to this field could be written as:
\beq\ket{\psi_I} = \sum\limits_{y=0}^{N_y-1} \sum\limits_{x=0}^{N_x-1} \sum\limits_{d=0}^{N_d-1}  u(x,y,d)\ket {y,x,d}\eeq
If we want to apply a remeshing along both directions $x$ and $y$, we first obtain the following ``duplicated'' state:
\beq\ket{\psi_{D}} = \sum\limits_{y=0}^{2N_y-1} \sum\limits_{x=0}^{2N_x-1} \sum\limits_{d=0}^{N_d-1}  u(x//2,y//2,d)\ket {y,x,d}\eeq
The reordering circuit is similar to the one seen before, but we also have to put the $q_d$ DOF qubits at the end, using swaps:

\begin{figure}[h!] 
    \centering
    \includegraphics[height=8.4em]{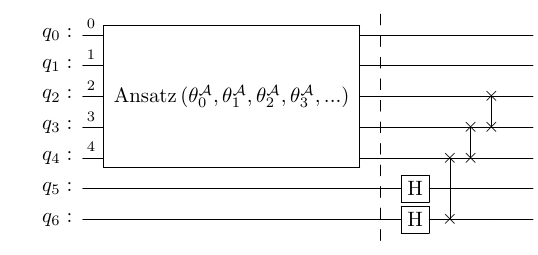}
    
    \caption{Circuit generating $\ket{\psi_D}$ in the $2$D vectorial case.}
\end{figure}

Notice that the need for swaps disappears in real quantum circuit execution, by simply using ansätze that don't use the wires that will be used for remeshing. For instance, in the case of the $5$-qubit to $7$-qubit remeshing of our $2$D plate, the first ansatz would use qubits $0,1,3,4,6$, while the second ansatz would use all seven. Moreover, by manipulating the endianness of the binary representation of the coordinates (writing the state as $\sum u(x,y,d)\ket{\bar y, d, x}$), we can achieve the 2D remeshing without swaps or temporarily unused wires ---see Fig. \ref{fig:ansatz2D_remeshed_noswap}.

The construction in 3D is identical to the one in 2D with one more swap ladder, since we have to reach the state $\ket{z+y+x+d}$ after duplication - see Appendix \ref{sec:casc:3D}.

\ifprlstyle
\begin{acknowledgments}
\fi

\head*{Acknowledgments} 
The authors would like to thank Paris Region for the support of Aqadoc project. This work is part of OECQ project which is financed by the French State as part of France 2030, financed by the European Union - Next Generation EU as part of the France Relance plan.
The authors would like to thank Patrick Sinnott and Nicolas Maring for useful discussions and Fabrice Debbasch for multiple technical advices.

\ifprlstyle
\end{acknowledgments}
\fi

\ifprlstyle
\else
\bibliographystyle{quantum}
\fi

\bibliography{main}

\begin{thebibliography}{10}

\bibitem{griffith1921vi}
Alan~Arnold Griffith.
\newblock ``Vi. the phenomena of rupture and flow in solids''.
\newblock \href{https://dx.doi.org/10.1098/rsta.1921.0006}{Philosophical
  transactions of the royal society of london. Series A, containing papers of a
  mathematical or physical character {\bf 221}, 163--198}~(1921).

\bibitem{Coussy2004}
Olivier Coussy.
\newblock ``Poromechanics''.
\newblock \href{https://dx.doi.org/10.1002/0470092718}{John Wiley \& Sons}.
  ~(2003).

\bibitem{Ramberg1943}
Walter Ramberg and William~R Osgood.
\newblock ``Description of stress-strain curves by three parameters''.
\newblock Technical report.
\newblock NASA~(1943).

\bibitem{Fontana25}
I.~Fontana, G.~Bacquaert, D.A. Di~Pietro, and K.~Kazymyrenko.
\newblock ``Hyperelastic nature of the hoek–brown criterion''.
\newblock \href{https://dx.doi.org/10.1016/j.euromechsol.2025.105782}{European
  Journal of Mechanics - A/Solids {\bf 115}, 105782}~(2026).

\bibitem{love2013}
A.E.H. Love.
\newblock ``A treatise on the mathematical theory of elasticity''.
\newblock Cambridge University Press. ~(2013).

\bibitem{TJRHughes}
Thomas~JR Hughes.
\newblock ``The finite element method: linear static and dynamic finite element
  analysis''.
\newblock Prentice-Hall. ~(1987).

\bibitem{Rüde}
Björn Gmeiner, Markus Huber, Lorenz John, Ulrich Rüde, and Barbara Wohlmuth.
\newblock ``A quantitative performance analysis for stokes solvers at the
  extreme scale''~(2015).
\newblock  \href{http://arxiv.org/abs/1511.02134}{arXiv:1511.02134}.

\bibitem{harrowQuantumAlgorithmSolving2009}
Aram~W. Harrow, Avinatan Hassidim, and Seth Lloyd.
\newblock ``Quantum algorithm for solving linear systems of equations''.
\newblock \href{https://dx.doi.org/10.1103/PhysRevLett.103.150502}{Physical
  Review Letters {\bf 103}, 150502}~(2009).

\bibitem{childsQuantumAlgorithmSystems2017a}
Andrew~M. Childs, Robin Kothari, and Rolando~D. Somma.
\newblock ``Quantum {Algorithm} for {Systems} of {Linear} {Equations} with
  {Exponentially} {Improved} {Dependence} on {Precision}''.
\newblock \href{https://dx.doi.org/10.1137/16M1087072}{SIAM Journal on
  Computing {\bf 46}, 1920--1950}~(2017).

\bibitem{caoQuantumAlgorithmCircuit2013a}
Yudong Cao, Anargyros Papageorgiou, Iasonas Petras, Joseph Traub, and Sabre
  Kais.
\newblock ``Quantum algorithm and circuit design solving the {Poisson}
  equation''.
\newblock \href{https://dx.doi.org/10.1088/1367-2630/15/1/013021}{New Journal
  of Physics {\bf 15}, 013021}~(2013).

\bibitem{berryHighorderQuantumAlgorithm2014}
Dominic~W. Berry.
\newblock ``High-order quantum algorithm for solving linear differential
  equations''.
\newblock \href{https://dx.doi.org/10.1088/1751-8113/47/10/105301}{Journal of
  Physics A: Mathematical and Theoretical {\bf 47}, 105301}~(2014).

\bibitem{bravo-prietoVariationalQuantumLinear2023a}
Carlos Bravo-Prieto, Ryan LaRose, M.~Cerezo, Yigit Subasi, Lukasz Cincio, and
  Patrick~J. Coles.
\newblock ``Variational {Quantum} {Linear} {Solver}''.
\newblock \href{https://dx.doi.org/10.22331/q-2023-11-22-1188}{Quantum {\bf 7},
  1188}~(2023).

\bibitem{demirdjianVariationalQuantumSolutions2022}
Reuben Demirdjian, Daniel Gunlycke, Carolyn~A. Reynolds, James~D. Doyle, and
  Sergio Tafur.
\newblock ``Variational quantum solutions to the advection–diffusion equation
  for applications in fluid dynamics''.
\newblock \href{https://dx.doi.org/10.1007/s11128-022-03667-7}{Quantum
  Information Processing{\bf 21}}~(2022).

\bibitem{huQuantumCircuitsPartial2024}
Junpeng Hu, Shi Jin, Nana Liu, and Lei Zhang.
\newblock ``Quantum {Circuits} for partial differential equations via
  {Schrödingerisation}''.
\newblock \href{https://dx.doi.org/10.22331/q-2024-12-12-1563}{Quantum {\bf 8},
  1563}~(2024).

\bibitem{lindenQuantumVsClassical2020}
Noah Linden, Ashley Montanaro, and Changpeng Shao.
\newblock ``Quantum vs. classical algorithms for solving the heat equation''.
\newblock \href{https://dx.doi.org/10.1007/s00220-022-04442-6}{Communications
  in Mathematical Physics {\bf 395}, 601--641}~(2022).

\bibitem{SurveyHHL}
Xiaonan Liu, Haoshan Xie, Zhengyu Liu, and Chenyan Zhao.
\newblock ``Survey on the improvement and application of hhl algorithm''.
\newblock \href{https://dx.doi.org/10.1088/1742-6596/2333/1/012023}{Journal of
  Physics: Conference Series {\bf 2333}, 012023}~(2022).

\bibitem{lloydQuantumAlgorithmNonlinear2020}
Seth Lloyd, Giacomo De~Palma, Can Gokler, Bobak Kiani, Zi-Wen Liu, Milad
  Marvian, Felix Tennie, and Tim Palmer.
\newblock ``Quantum algorithm for nonlinear differential equations''~(2020).
\newblock  \href{http://arxiv.org/abs/2011.06571}{arXiv:2011.06571}.

\bibitem{liuVariationalQuantumAlgorithm2021a}
Hailing Liu, Yusen Wu, Linchun Wan, Shijie Pan, Sujuan Qin, Fei Gao, and
  Qiaoyan Wen.
\newblock ``Variational {Quantum} algorithm for {Poisson} equation''.
\newblock \href{https://dx.doi.org/10.1103/PhysRevA.104.022418}{Physical Review
  A {\bf 104}, 022418}~(2021).

\bibitem{Sato}
Yuki Sato, Ruho Kondo, Satoshi Koide, Hideki Takamatsu, and Nobuyuki Imoto.
\newblock ``Variational quantum algorithm based on the minimum potential energy
  for solving the {Poisson} equation''.
\newblock \href{https://dx.doi.org/10.1103/PhysRevA.104.052409}{Physical Review
  A {\bf 104}, 052409}~(2021).

\bibitem{kroviImprovedQuantumAlgorithms2023}
Hari Krovi.
\newblock ``Improved quantum algorithms for linear and nonlinear differential
  equations''.
\newblock \href{https://dx.doi.org/10.22331/q-2023-02-02-913}{Quantum {\bf 7},
  913}~(2023).

\bibitem{McClean2018}
Jarrod~R. McClean, Sergio Boixo, Vadim~N. Smelyanskiy, Ryan Babbush, and
  Hartmut Neven.
\newblock ``Barren plateaus in quantum neural network training landscapes''.
\newblock \href{https://dx.doi.org/10.1038/s41467-018-07090-4}{Nature
  Communications {\bf 9}, 4812}~(2018).

\bibitem{Cerezo2021}
M.~Cerezo, Akira Sone, Tyler Volkoff, Lukasz Cincio, and Patrick~J. Coles.
\newblock ``Cost function dependent barren plateaus in shallow parametrized
  quantum circuits''.
\newblock \href{https://dx.doi.org/10.1038/s41467-021-21728-w}{Nature
  Communications{\bf 12}}~(2021).

\bibitem{maring2024versatile}
Nicolas Maring, Andreas Fyrillas, Mathias Pont, Edouard Ivanov, Petr Stepanov,
  Nico Margaria, William Hease, Anton Pishchagin, Aristide Lema{\^\i}tre,
  Isabelle Sagnes, et~al.
\newblock ``A versatile single-photon-based quantum computing platform''.
\newblock \href{https://dx.doi.org/10.1038/s41566-024-01403-4}{Nature Photonics
  {\bf 18}, 603--609}~(2024).

\bibitem{irwin1957}
G.~R. Irwin.
\newblock ``{Analysis of Stresses and Strains Near the End of a Crack
  Traversing a Plate}''.
\newblock \href{https://dx.doi.org/10.1115/1.4011547}{Journal of Applied
  Mechanics {\bf 24}, 361--364}~(1957).

\bibitem{westergaard39}
H.~M. {Westergaard}.
\newblock ``{Bearing Pressures and Cracks: Bearing Pressures Through a Slightly
  Waved Surface or Through a Nearly Flat Part of a Cylinder, and Related
  Problems of Cracks}''.
\newblock \href{https://dx.doi.org/10.1115/1.4008919}{Journal of Applied
  Mechanics {\bf 6}, A49--A53}~(1939).

\bibitem{Murakami}
Y.~Murakami.
\newblock ``Stress intensity factors handbook''.
\newblock \href{https://dx.doi.org/10.1115/1.2900983}{Pergamon}. ~(1987).
\newblock 1st ed. edition.

\bibitem{BrownSrawley}
William~F. Brown and John~E. Srawley.
\newblock ``Plane strain crack toughness testing of high strength metallic
  materials''.
\newblock \href{https://dx.doi.org/10.1520/stp44663s}{Page 1–129}.
\newblock ASTM International100 Barr Harbor Drive, PO Box C700, West
  Conshohocken, PA 19428-2959. ~(1966).

\bibitem{Zyl23}
Julien Zylberman and Fabrice Debbasch.
\newblock ``{Efficient quantum state preparation with Walsh series}''.
\newblock \href{https://dx.doi.org/10.1103/physreva.109.042401}{Physical Review
  A{\bf 109}}~(2024).

\bibitem{Ern2004}
Alexandre Ern and Jean-Luc Guermond.
\newblock ``Finite element interpolation''.
\newblock In Theory and Practice of Finite Elements.
\newblock \href{https://dx.doi.org/10.1007/978-1-4757-4355-5_1}{Volume 159 of
  Applied Mathematical Sciences, pages 3--80}.
\newblock Springer~(2004).

\bibitem{lysaght2024quantum}
Liam Lysaght, Timoth{\'e}e Goubault, Patrick Sinnott, Shane Mansfield, and
  Pierre-Emmanuel Emeriau.
\newblock ``Quantum circuit compression using qubit logic on qudits''~(2024).
\newblock  \href{http://arxiv.org/abs/2411.03878}{arXiv:2411.03878}.

\bibitem{somaschi2016near}
Niccolo Somaschi, Valerian Giesz, Lorenzo De~Santis, JC~Loredo, Marcelo~P
  Almeida, Gaston Hornecker, S~Luca Portalupi, Thomas Grange, Carlos Anton,
  Justin Demory, et~al.
\newblock ``Near-optimal single-photon sources in the solid state''.
\newblock \href{https://dx.doi.org/10.1038/nphoton.2016.23}{Nature Photonics
  {\bf 10}, 340--345}~(2016).

\bibitem{fyrillas2024high}
Andreas Fyrillas, Olivier Faure, Nicolas Maring, Jean Senellart, and Nadia
  Belabas.
\newblock ``High-fidelity quantum information processing with machine
  learning-characterized photonic circuits''.
\newblock In Quantum 2.0.
\newblock \href{https://dx.doi.org/10.1364/QUANTUM.2024.QW4A.1}{Pages QW4A--1}.
\newblock Optica Publishing Group~(2024).

\bibitem{Lame1833}
Gabriel Lam{\'e} and Emile Clapeyron.
\newblock ``{M{\'e}moire sur l'{\'e}quilibre int{\'e}rieur des corps solides
  homog{\`e}nes}''.
\newblock M{\'e}moires de l'Acad{\'e}mie des sciences de l'Institut de
  France~(1833).

\bibitem{griffith1921}
Alan~Arnold Griffith and Geoffrey~Ingram Taylor.
\newblock ``Vi. the phenomena of rupture and flow in solids''.
\newblock \href{https://dx.doi.org/10.1098/rsta.1921.0006}{Philosophical
  Transactions of the Royal Society of London. Series A, Containing Papers of a
  Mathematical or Physical Character {\bf 221}, 163--198}~(1921).

\bibitem{dhatt2012}
Gouri Dhatt, Gilbert Touzot, and Emmanuel Lefrancois.
\newblock ``Finite element method''.
\newblock \href{https://dx.doi.org/10.1002/9781118569764.fmatter}{John Wiley \&
  Sons}. ~(2012).

\bibitem{r3.01.01}
Documentation Code\_Aster EDF R\& D: R3.01.01.
\newblock ``Fonctions de forme et points d’int{\'{e}}gration des
  {\'{e}}l{\'{e}}ments finis''.
\newblock
  url:~\url{https://code-aster.org/V2/doc/default/fr/man\_r/r3/r3.01.01.pdf}.

\bibitem{powellNEWUOASoftwareUnconstrained2006}
M.~J.~D. Powell.
\newblock ``The {{NEWUOA}} software for unconstrained optimization without
  derivatives''.
\newblock In G.~Di~Pillo and M.~Roma, editors, Large-{{Scale Nonlinear
  Optimization}}.
\newblock \href{https://dx.doi.org/10.1007/0-387-30065-1_16}{Pages 255--297}.
\newblock Springer US, Boston, MA~(2006).

\bibitem{heurtel2023strong}
Nicolas Heurtel, Shane Mansfield, Jean Senellart, and Beno{\^\i}t Valiron.
\newblock ``Strong simulation of linear optical processes''.
\newblock \href{https://dx.doi.org/10.1016/j.cpc.2023.108848}{Computer Physics
  Communications {\bf 291}, 108848}~(2023).

\bibitem{barsoum1976}
Roshdy~S Barsoum.
\newblock ``On the use of isoparametric finite elements in linear fracture
  mechanics''.
\newblock \href{https://dx.doi.org/10.1002/nme.1620100103}{International
  Journal for Numerical Methods in Engineering {\bf 10}, 25--37}~(1976).

\end{thebibliography}

\newpage
\onecolumngrid%
\appendix

\section{Extensions of quantum remeshing}
In the main part of this work, the quantum remeshing technique was introduced and detailed for one- and two-dimensional cases. In this appendix, we discuss several complementary topics related to quantum remeshing. We first focus on the extension to three dimensions, then introduce variants of quantum remeshing that avoid the use of swap gates, which may be advantageous for certain QPU hardware architectures. Finally, we provide several suggestions aimed at mitigating the field relaxation issues observed in our simulations.

\subsection{Construction in 3D}\label{sec:casc:3D}

We illustrate here how the quantum remeshing technique could be extended into $3D$ case.
A reasonable minimal $3D$ example can be constructed from the initial system with $8$-qubits. It corresponds to a $2$-qubit discretization in each of the three directions, and four local DoFs ($d=4$) that are supposed to be encoded with two complementary qubits. For instance, we could refer to the thermo-mechanical coupling problem, where the unknowns are the three component vector-displacements and the temperature field. In $3D$, at each remeshing step, three qubits must be added to represent the homogeneous refinement in each spatial direction.
\begin{figure}[H] 
    \centering

    \subfloat[The initially optimized $8$-qubit ansatz, with 3 extra qubits.\label{fig:ansatz3DVEC}]{\makebox[0.45\textwidth]{\includegraphics[height=9.5em]{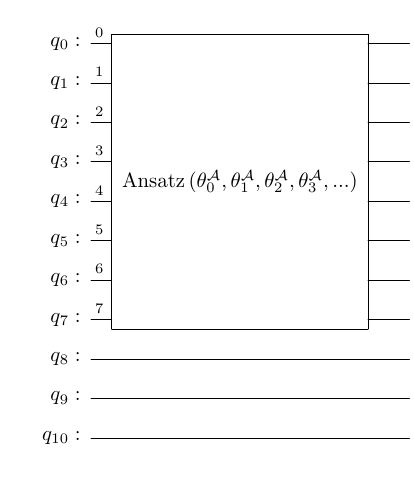}}}
    \qquad

    \subfloat[Field duplication and reordering within quantum circuit.\label{fig:ansatz3DVEC+wire}]{\makebox[0.45\textwidth]{\includegraphics[height=9.5em]{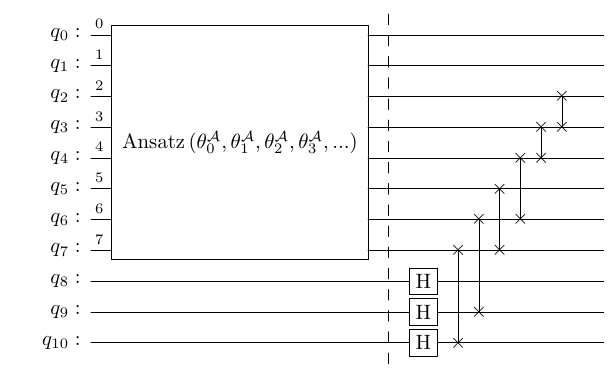}}}

    \caption{Construction of the remeshed $3D$ ansatz, illustrating the transition:$\ket {xyzd}\otimes\ket{000}\mapsto \ket{x}\ket+\ket y \ket+ \ket{z}\ket+\ket d$.}
\end{figure}

In the reordering circuit, the first two swaps place the $d$ components at the end, the second two swaps place the two $x$ values before a $\ket+$, and the last two swaps place the two $y$ values before a $\ket+$. 

\begin{figure}[htbp]
    \centering
    \includegraphics[width=.5\linewidth]{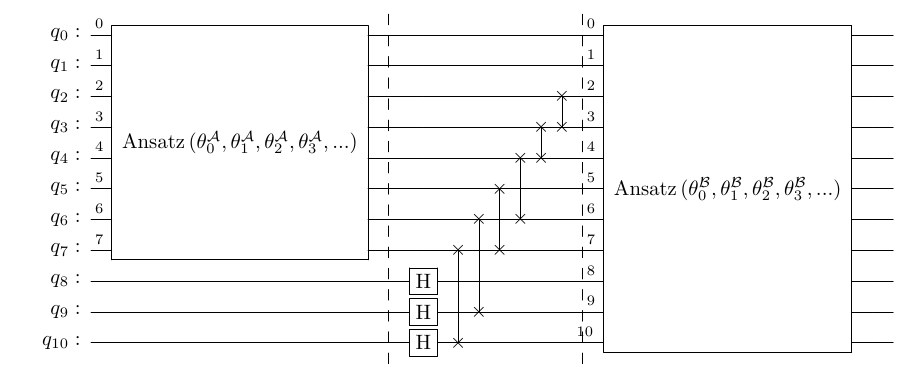}
    \caption{Circuit $\calC$ to be optimized towards $T_{11}$}
    \label{fig:ansatz3D_remeshed}
\end{figure}

\begin{figure}[htbp]
\centering
\subfloat[The deformation level on the $13$ qubits cascaded quantum solution for the $2D$ plate.\label{fig:2Dmesh:quantum13q}]{\includegraphics[width=0.325\textwidth]{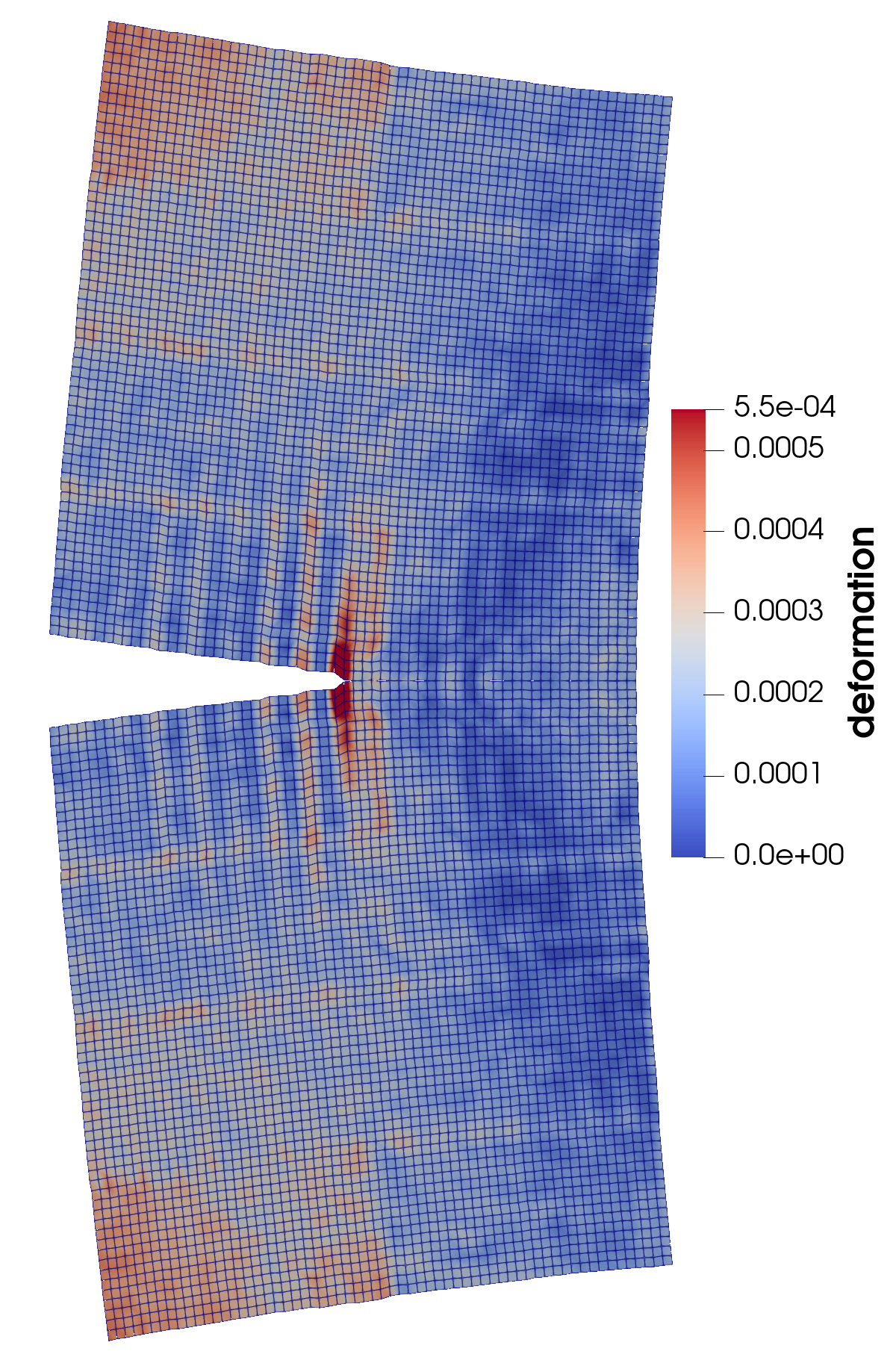}}\hfill
\subfloat[The same as (a) for the next step remeshing $15$ qubits. Field relaxation problem is clearly amplified.\label{fig:2Dmesh:quantum15q}]{\includegraphics[width=0.27\textwidth]{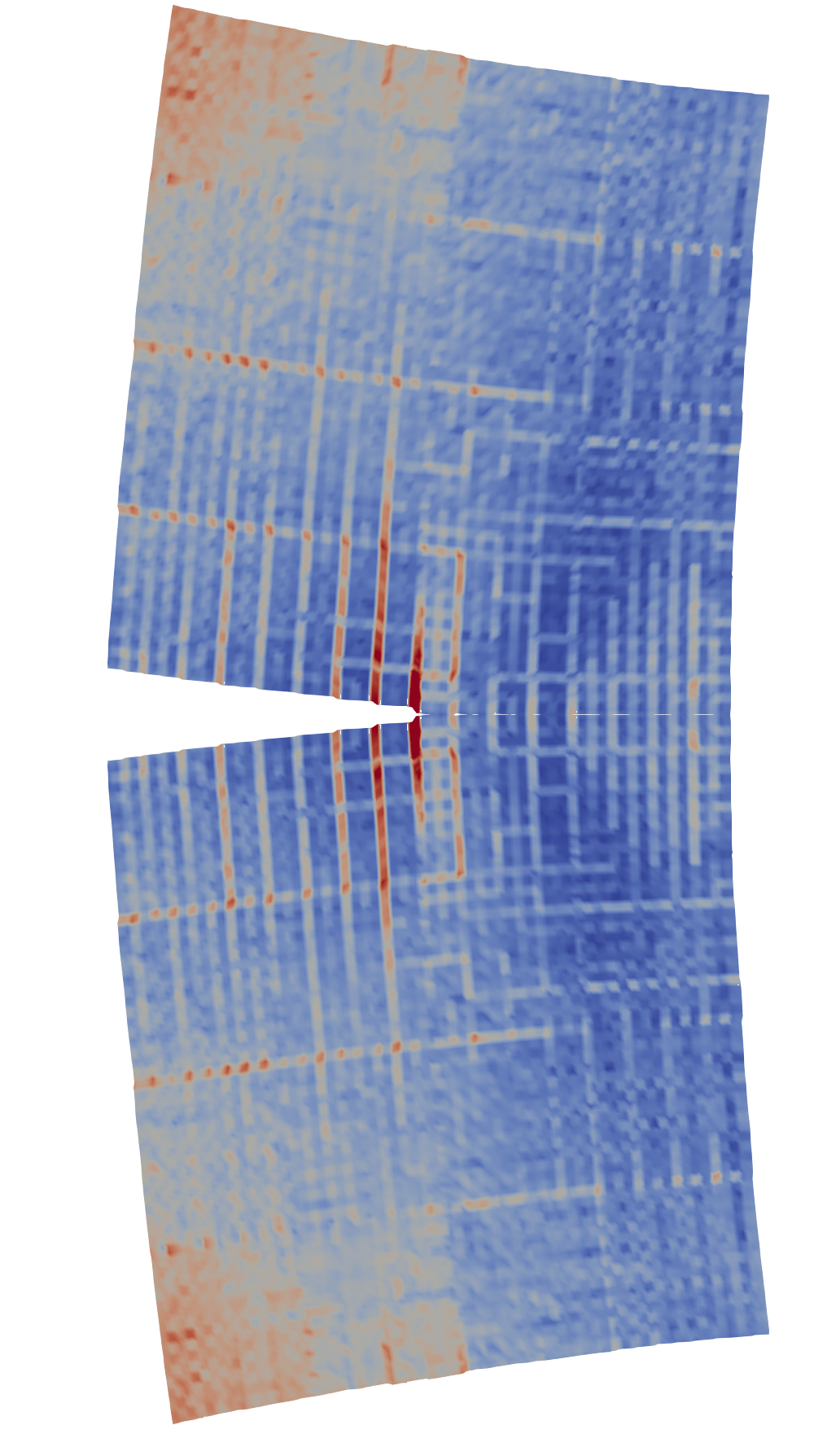}}\hfill
\hfill
\subfloat[The cascade remeshing applied to a 3D shape.\label{fig:3Dmesh:numbered}]{\includegraphics[width=0.33\textwidth]{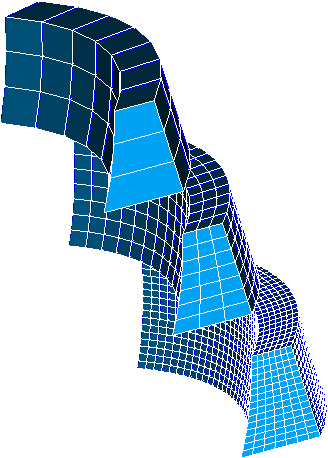}}
\caption{Usual discretized geometries.} \label{fig:1}
\end{figure}
\FloatBarrier

\subsection{Construction without swaps}\label{sec:2D:noswap}
While photonic QPUs offer a relatively flexible implementation of SWAP gates due to their adaptable connectivity, this is not the case for all quantum architectures. In particular, superconducting QPUs often suffer from limited qubit connectivity, where qubits can only interact with their nearest neighbors. As a result, implementing a SWAP gate between distant qubits requires decomposing it into multiple double-qubit operations, significantly increasing circuit depth and error rates. To address this challenge, we propose a swapless variant of quantum remeshing, which avoids the need for costly SWAP operations and is better suited for architectures with constrained connectivity.

In this section, we use the index $\cdot_b$ to indicate that a number is written in binary. A number can be written in binary either in little-endian ($34_b = 100010$) or in big-endian ($\bar {34}_b = 010001$). We can go from one endianness to another by mirroring the representation.

In $2D$, it is possible to add qubits in a way that does not significantly affect the nodal enumeration and allows the creation of observables, without the need for swaps. We should simply have $y$ represented by its {big-endian} binary representation $\bar y_b$, followed by $d$ as usual, followed by $x$ represented as before by its {little-endian} binary representation $x_b$. The two remeshing qubits have to be added at the top and bottom of the ansatz:
\begin{figure}[htbp]
    \centering
    \includegraphics[width=.5\linewidth]{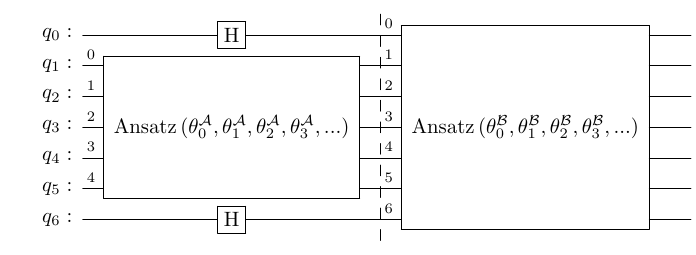}
    \caption{Circuit without swaps}
    \label{fig:ansatz2D_remeshed_noswap}
\end{figure}
The ansatz state at the dotted barrier in Fig. \ref{fig:ansatz2D_remeshed_noswap} is:
\beq\begin{aligned}\ket{\psi_{NOSWAP}}&= \sum\limits_{y=0}^{N_y-1} \sum\limits_{x=0}^{N_x-1} \sum\limits_{d=0}^{N_d-1}  \ddepl_{\calA,4}(x,y,d)\ket {+,\bar y_b,d,x_b,+}\\
&= \sum\limits_{y=0}^{2N_y-1} \sum\limits_{x=0}^{2N_x-1} \sum\limits_{d=0}^{N_d-1}  \ddepl_{\calA,4}(x//2,y//2,d)\ket {\bar y_b,d,x_b}\end{aligned}\eeq

Another possibility is to design the ansätze for lower number of qubits so that they do not use some of their internal qubits. This way, they are available for remeshing by simply applying $H$ to them, like in Fig. \ref{fig:ansatz2D_remeshed_noswap}, without any need for endianness modifications.

However, the advantage of these methods that output states in the standard $\ket {yxd}$ qubits order  %
is that the observable construction follows well-established rules. That observable creation process could be complexified:
\bite\item The swapless method shown above requires a different handling of $x$ and $y$. 
\item A swapless method, without any endianness changes, results, after a few rounds of remeshing, in basis states that have to be interpreted as $\ket {\bar x_0\bar y_0dx_1y_1x_2y_2x_3y_3\dots}$. The advantage here is that at each remeshing step, the hardware-efficient ansatz directly entangle the newly added qubits with one-another, as explained in Sec. \ref{sec:casc:ans}. 
\eite

We note that, while in $2D$ it is possible to absorb all swap gates into initial smart qubits reordering \ref{sec:2D:noswap}, in $3D$ it is not possible to iteratively construct in qiskit a swapless circuit without requiring unused internal qubits. 
However, as in $2D$, we could still design an ansatz that does not use some of its internal qubits, which stay to $\ket0$ until the duplication applies to them.

\subsection{Ansatz specification improvement}\label{sec:casc:ans}
No matter the underlying equation that is treated, the optimization of the ansatz structure is a crucial question. For instance, in order to limit the number of swaps, we could apply the procedures discussed in Sec. \ref{sec:2D:noswap}. Another important example is linked to the field relaxation problem. Indeed, when applying a remeshing, we transform a smooth function into a piecewise constant function. This leads to a significant increase in error for most physically relevant observables, which typically exhibit continuity and are highly sensitive to disruptions in spatial derivatives. For instance, in fracture mechanics,
mechanical strain, defined as the gradient of the displacement field, directly contributes to the calculation of elastic energy. The so-created discontinuities in the mechanical strain represent an extremely high local elastic potential, and hence account for a large part of the energy modification.

The newly added nodes contributes the most for local variations - regardless of the endianness of the problem. Given that errors after such a relaxation are mostly local, it would be advised to ensure a reasonable entanglement of these nodes with the other ones, but also between them.

As such, given that the usual hardware efficient ansatz mostly entangles close qubits together, a simple proposition would be to place, using swaps and endianness manipulations as in Sec. \ref{sec:2D:noswap}, all these qubits together within the ansatz. It would still be possible to replace them at their right spot afterwards, so that reading and applying observables would not be modified. A pre-ansatz GHZ state-based entanglement procedure between these $D$ newly-added qubits could also be of use.

\begin{figure}[htbp]
    \centering
    \includegraphics[width=.83\linewidth]{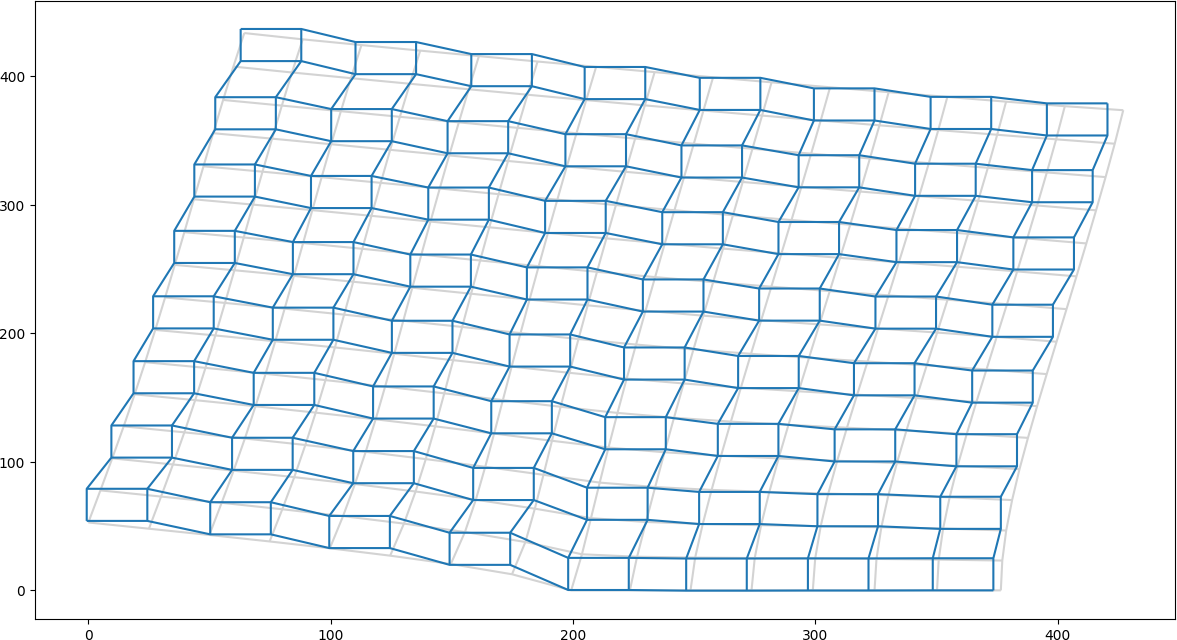}
    \caption{\centering A field directly after remeshing (in blue), compared with the optimal values (in grey). The clusters of 4 identical displacement values are visible, especially in the bottom left area.}
    \label{fig:relaxation}
\end{figure}
\newpage

\subsection{Rescaling considerations}\label{sec:resc}

In the usual version of a remeshing, we increase the number of elements while keeping the initial nodes at the same positions.
Quantum remeshing however duplicates the number of nodes, resulting in a new non-conformal mesh: the nodes are created orderly from the extremal nodes, and the former nodes' positions are shifted in order to respect mesh regularity.  We illustrate this subtle point on a simple $1$D example.
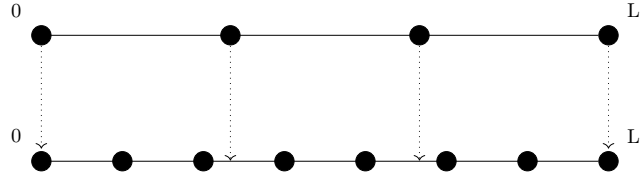
\begin{figure}[htbp]
\centering
\resizebox{.48\textwidth}{!}{%
\begin{tikzpicture}
   \node (AL) at (0.4,0.4) {L};
   \node (A) [fill=black, shape=circle, draw=black]at (0, 0) {};
   \draw[dotted,->] (0, 0) --(0, -1.8);
   \node (B) [fill=black, shape=circle, draw=black]at (-3, 0) {};
   \draw[dotted,->] (-3, 0) --(-3, -1.99);
   \node (C) [fill=black, shape=circle, draw=black] at (-6, 0) {};
   \draw[dotted,->] (-6, 0) --(-6, -1.99);
   \node (D) [fill=black, shape=circle, draw=black] at (-9, 0) {};
   \draw[dotted,->] (-9, 0) --(-9, -1.8);
   \node (DL) at (-9.4,0.4) {0};
   \draw (A) edge (-9, 0);
   
   \node (A2L) at (0.4,-1.6) {L};
   \node (A2) [fill=black, shape=circle, draw=black]at (0, -2) {};
   \node (A1) [fill=black, shape=circle, draw=black]at (-9/7, -2) {};
   \node (B2) [fill=black, shape=circle, draw=black]at (-18/7, -2) {};
   \node (B2) [fill=black, shape=circle, draw=black]at (-27/7, -2) {};
   
   \node (D2) [fill=black, shape=circle, draw=black]at (-9+27/7, -2) {};
   \node (D2) [fill=black, shape=circle, draw=black]at (-9+18/7, -2) {};
   \node (D2) [fill=black, shape=circle, draw=black]at (-9+9/7, -2) {};
   \node (D2) [fill=black, shape=circle, draw=black]at (-9, -2) {};
   \node (D2L) at (-9.4,-1.6) {0};
   \draw (A2) edge (-9, -2);
\end{tikzpicture}

}%
\caption{$1$-D remeshing from $2$ to $3$ qubits.}
\label{fig:1D:rescaling}
\end{figure}

Consider a $1D$, $4$-nodes discretized field ($N = 4$) corresponding to a $2$-qubit system. The third point ($n=3$) is associated with the basis state $\ket {11}$ and has an amplitude that will be noted as $T_{2,3}$ ($T_2$ being the entire field and $T_{2,3}$ its fourth component, its first being $T_{2,0}$).
We can note its position as $p_{2,3} = L$, and the position of the first point as $p_{2,0} = 0$. That way, the $n$-th point has position $p_{2,n} = \frac{nL}{3}$.
We will now apply a remeshing step. Consider the point associated to $\ket {000}$. Since we are still operating between positions $0$ and $L$, its position is still at $p_{3,0} = 0$. The next point is associated to $\ket {001}$. Its position is at $p_{3,1} = L/7$. Notice that $p_{3,0} \leq p_{2,0}\leq p_{3,1}$, meaning that it lies inside the element bounded by its two nodal successors.

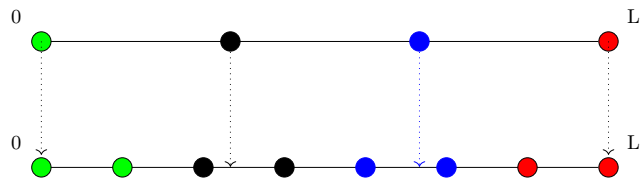
\begin{figure}[htbp]
\centering
\resizebox{.48\textwidth}{!}{%
\begin{tikzpicture}
   \node (AL) at (0.4,0.4) {L};
   \node (A) [fill=red, shape=circle, draw=black]at (0, 0) {};
   \draw (A) edge (-9, 0);
   \draw[dotted,->] (0, 0) --(0, -1.8);
   \node (B) [fill=blue, shape=circle, draw=blue]at (-3, 0) {};
   \draw[dotted,draw=blue,->] (-3, 0) --(-3, -1.99);
   \node (C) [fill=black, shape=circle, draw=black] at (-6, 0) {};
   \draw[dotted,->] (-6, 0) --(-6, -1.99);
   \node (D) [fill=green, shape=circle, draw=black] at (-9, 0) {};
   \draw[dotted,->] (-9, 0) --(-9, -1.8);
   \node (DL) at (-9.4,0.4) {0};
   
   \node (A2L) at (0.4,-1.6) {L};
   \node (A2) [fill=red, shape=circle, draw=black]at (0, -2) {};
   \draw (A2) edge (-9, -2);
   \node (A1) [fill=red, shape=circle, draw=black]at (-9/7, -2) {};
   \node (B2) [fill=blue, shape=circle, draw=blue]at (-18/7, -2) {};
   \node (B2) [fill=blue, shape=circle, draw=blue]at (-27/7, -2) {};
   
   \node (D2) [fill=black, shape=circle, draw=black]at (-9+27/7, -2) {};
   \node (D2) [fill=black, shape=circle, draw=black]at (-9+18/7, -2) {};
   \node (D2) [fill=green, shape=circle, draw=black]at (-9+9/7, -2) {};
   \node (D2) [fill=green, shape=circle, draw=black]at (-9, -2) {};
   \node (D2L) at (-9.4,-1.6) {0};
\end{tikzpicture}

}%
\caption{Highlighting the node association.}
\label{fig:1D:rescaling:color}
\end{figure}
Consider the node associated to $\ket {10}$. Its position is at $p_{2,2} = 2L/3$. Its remeshed (successor) nodes are associated to ket $\ket {100}$ and $\ket{101}$. Notice that $p_{3,4} \leq p_{2,2}\leq p_{3,5}$, meaning that it is between its successor nodes as well.

More generally, consider the point associated to $\ket n$ in a $q$-qubit field. Its position is $p_{q,n} = \frac{nL}{2^{q}-1}$ and the position of its remeshed points are:
\begin{equation*}
    p_{q+1,2n} = \frac{2nL}{2^{q+1}-1} \quad\text{and}\quad  p_{q+1,2n+1} = \frac{(2n+1)L}{2^{q+1}-1},
\end{equation*}
Inner projection inequality is satisfied:
\beq p_{q+1,2n} \leq p_{q,n} \leq p_{q+1,2n+1} \eeq
Indeed,
\beq  p_{q+1,2n} = \frac{nL}{2^{q}-\hf} \leq \frac{nL}{2^{q}-1} = p_{q,n} \eeq
\beq\begin{aligned}\frac{p_{q+1,2n+1} - p_{q,n}}{L} =& \frac{2n+1}{2^{q+1}-1} - \frac{n}{2^{q}-1}\\
	=& \frac{2^{q}-n-1}{(2^{q+1}-1)(2^{q}-1)} \geq 0 \quad \\ 
    \forall n\in[0,2^{q}-1]\cap \mathbb N
\end{aligned}\eeq

Similarly, in $2$-D and $3$-D, the initial node is in the square (resp. the cube) formed by its successors.

\section{Decomposition of the stiffness matrix}
In this section, we present the complete procedure for decomposing the stiffness matrix into elementary quantumly computable observables, which scale logarithmically with the system size.

\subsection{Elementary connecting links \texorpdfstring{$\bbK_{ij}$}{K\_ij}}\label{sec:appendix:Kij}
The computation of the stiffness matrix $\bbK$ constitutes the main task of the finite element discretization of the Navier-Cauchy mechanical problem \eqref{eq:Navier-Cauchy}. For the regular meshes the tensor product decomposition  \eqref{eq:K_decomp} allows one to present it in a compact form. Any elementary connecting link $\bbK_{ij}$ introduced in \eqref{eq:K_decomp} can be evaluated using its integral expression given in \eqref{eq:Kij_all}:

\begin{equation}\label{eq:dNdN_aa}
    \mathbb{K}_{aa}=\mathbb{K}_{cc}=
    \begin{pmatrix}
    1/2 - 2\nu/3 & 1/8\\ 
    1/8 & 1/2 - 2\nu/3
  \end{pmatrix}; \quad
\end{equation}

\begin{equation} \label{eq:dNdN_bb}  
    \mathbb{K}_{bb}=\mathbb{K}_{dd}=
    \begin{pmatrix}
    1/2 - 2\nu/3 & -1/8\\ 
    -1/8 & 1/2 - 2\nu/3
  \end{pmatrix}; \quad  
\end{equation}

\begin{equation}\label{eq:dNdN_ab}
    \mathbb{K}_{ab}=\mathbb{K}_{cd}=
    \begin{pmatrix}
    -1/4 + \nu/6& -1/8+\nu/2\\ 
    1/8-\nu/2 & \nu/6
  \end{pmatrix}; \quad 
\end{equation}

\begin{equation}\label{eq:dNdN_ad}
    \mathbb{K}_{ad}= 
    \mathbb{K}_{cb}=
    \begin{pmatrix}
    \nu/6& 1/8 -\nu/2 \\ 
    -1/8 +\nu/2 & -1/4+\nu/6
  \end{pmatrix}; \quad
\end{equation}

\begin{equation} \label{eq:dNdN_ac} 
    \mathbb{K}_{ac}=
    \mathbb{K}_{ca}=
    \begin{pmatrix}
    -1/4 + \nu/3 & -1/8\\ 
    -1/8 & -1/4 + \nu/3
  \end{pmatrix}; \quad  
\end{equation}

\begin{equation} \label{eq:dNdN_bd} 
    \mathbb{K}_{bd}=
    \mathbb{K}_{db}=
    \begin{pmatrix}
    -1/4 + \nu/3 & 1/8\\ 
    1/8 & -1/4 + \nu/3
  \end{pmatrix}. 
\end{equation}
These link matrices are to be decomposed into the Pauli basis \eqref{eq:setS} prior to any quantum calculations:

\begin{align}
    \bbK_{aa} & = \frac{3+4\nu}{6}I + \frac{1}{8}X; \qquad
    \bbK_{bb}  =  \frac{3+4\nu}{6}I - \frac{1}{8}X;\\
    \bbK_{ab} & =  -\frac{3+4\nu}{24}I - \frac{1}{8}Z + \frac{4\nu-1}{8}iY; \\
    \bbK_{bc} & =  -\frac{3+4\nu}{24}I + \frac{1}{8}Z + \frac{4\nu-1}{8}iY;\\
    \bbK_{ac} & =  -\frac{3+4\nu}{12}I - \frac{1}{8}X; \quad 
    \bbK_{bd}  =  -\frac{3+4\nu}{12}I + \frac{1}{8}X.
\end{align}
The procedure could be extended for other relevant situations, such as the discretized scalar partial differential equations (PDE). For example, the tensor product decomposition for Poisson PDE can be trivially obtained from \eqref{eq:K_decomp} by replacing all connecting links $\bbK_{ij}$ by its scalar counterparts from the elementary stiffness matrix: $\bbK_{ij}\mapsto K_{ij}$. The finite elements discretization for the Poisson equation is generated by the following set of $K_{ij}$:
\begin{equation}
\begin{bmatrix}
K_{aa} & K_{ab} & K_{ac} & K_{ad} \\
K_{ba} & K_{bb} & K_{bc} & K_{bd} \\
K_{ca} & K_{cb} & K_{cc} & K_{cd} \\
K_{da} & K_{db} & K_{dc} & K_{dd} 
\end{bmatrix} =
\begin{bmatrix}
4 & -1 & -2 & -1 \\
-1 & 4 & -1 & -2 \\
-2 & -1 & 4 & -1 \\
-1 & -2 & -1 & 4 \\
\end{bmatrix}    
\end{equation}
The same principle applies for the nearest neighbors finite difference scheme (FDM):
\begin{equation}
\begin{bmatrix}
K_{aa} & K_{ab} & K_{ac} & K_{ad} \\
K_{ba} & K_{bb} & K_{bc} & K_{bd} \\
K_{ca} & K_{cb} & K_{cc} & K_{cd} \\
K_{da} & K_{db} & K_{dc} & K_{dd} 
\end{bmatrix} =
\begin{bmatrix}
2 & -1 & 0 & -1 \\
-1 & 2 & -1 & 0 \\
0 & -1 & 2 & -1 \\
-1 & 0 & -1 & 2 \\
\end{bmatrix}    
\end{equation}

\subsection{Compact form of \texorpdfstring{$\bbK{}{}$}{K}}

Using the system's symmetry, we rewrite the full rigidity matrix \eqref{eq:K_decomp} into a compact form containing only $8$ independent contributions that are proportional to the elementary connecting link, defined in Appendix \ref{sec:appendix:Kij}:
\begin{equation}
\begin{aligned}\label{eq:decomposed_rigidity_K}
\mathbb{K} = & \left[(I^{\otimes n_y} - p_{-}^{\otimes n_y})\otimes(I^{\otimes n_x} - p_{-}^{\otimes n_x}) + (I^{\otimes n_y} - p_{+}^{\otimes n_y})\otimes(I^{\otimes n_x} - p_{+}^{\otimes n_x}) \right]\otimes \mathbb{K}_{aa} \: + \\ 
& \left[(I^{\otimes n_y} - p_{-}^{\otimes n_y})\otimes(I^{\otimes n_x} - p_{+}^{\otimes n_x}) + (I^{\otimes n_y} - p_{+}^{\otimes n_y})\otimes(I^{\otimes n_x} - p_{-}^{\otimes n_x}) \right]\otimes \mathbb{K}_{bb} \: + \\
& \left[(I^{\otimes n_y} - p_{-}^{\otimes n_y})\otimes \mathbb{T}(n_x) + (I^{\otimes n_y} - p_{+}^{\otimes n_y})\otimes \mathbb{T}^\dagger(n_x) \right]\otimes \mathbb{K}_{ab} \: + \\
& \left[\mathbb{T}(n_y)\otimes(I^{\otimes n_x} - p_{+}^{\otimes n_x}) + \mathbb{T}^\dagger(n_y)\otimes(I^{\otimes n_x} - p_{-}^{\otimes n_x}) \right]\otimes \mathbb{K}_{bc} \: + \\
& \left[\mathbb{T}(n_y)\otimes \mathbb{T}(n_x) + \mathbb{T}^\dagger(n_y)\otimes \mathbb{T}^\dagger(n_x) \right]\otimes \mathbb{K}_{ac} \: + \\
& \left[\mathbb{T}(n_y)\otimes \mathbb{T}^\dagger(n_x) + \mathbb{T}^\dagger(n_y)\otimes \mathbb{T}(n_x) \right]\otimes \mathbb{K}_{bd} \: + \\
& \left[\mathbb{T}(n_y)\otimes(I^{\otimes n_x} - p_{-}^{\otimes n_x}) + \mathbb{T}^\dagger(n_y)\otimes(I^{\otimes n_x} - p_{+}^{\otimes n_x}) \right]\otimes \mathbb{K}_{bc}^\dagger \: + \\
& \left[(I^{\otimes n_y} - p_{-}^{\otimes n_y})\otimes \mathbb{T}^\dagger(n_x) + (I^{\otimes n_y} - p_{+}^{\otimes n_y})\otimes \mathbb{T}(n_x) \right]\otimes \mathbb{K}_{ab}^\dagger.
\end{aligned}
\end{equation}
As shown in Appendix \ref{sec:appendix:Kij} all the connecting $2\times2$ matrices $\bbK_{ij}$ (\eqref{eq:Kij_all}) can be 
decomposed in the Pauli basis, with the help of linear coefficients dependent exclusively on the material's Poisson ratio $\nu$. The non-diagonal matrices used in the $\bbK$ decomposition are them-self represented with help of the elementary $2\times2$ matrices set:
\begin{equation*}
    \mathbb{T}(n_\alpha)=\sum\limits_{k=0}^{n_\alpha-1} I ^{\otimes k} \otimes \sigma_+ \otimes \sigma_-^{\otimes n_\alpha-k-1}.
\end{equation*}
As arbitrary linear combinations of $n$-times tensor products of $2\times 2$ matrices ($\mathcal{M}_{2}(\mathbb R)$ space) generate the $\mathcal{M}_{2^n}(\mathbb R)$ linear vector space, we can decompose $\mathbb{K} \in\mathcal{M}_{2^n}(\mathbb R)$ in the linear span of the polynomial tensor product of its four basis elements $\{p_\pm,\sigma_\pm\}$, which scales exponentially. The number of spanning terms in most general case of dense matrix scales as the corresponding vector space dimension, i.e. as $2^n\times 2^n$. 
One of our important claim is that the decomposition of $\mathbb{K}$  contains only
$O(n_x n_y)$-terms in a specific linear span generated by five elements
$\mathcal{S}=\{p_\pm,\sigma_\pm, I\}$. Even if it contains the non hermitian $\sigma_\pm$ operators, we prove in the Appendix \ref{sec:appendix:sigma_pm} that all these contributions can be regrouped in a are quantumly-computable observables.

\subsection{Diagonalization of \texorpdfstring{$\sigma_\pm$}{sigma+-} polynomials}\label{sec:appendix:sigma_pm}

\begin{figure}[htbp]
\centering
\subfloat[A $5$-qubit $\mathbb{B}_H(5)$ circuit for $\sigma_+^{\otimes 5}+\sigma_-^{\otimes 5}$.\label{fig:app:cnot_5q}]{%
\includegraphics[width=0.5\linewidth]{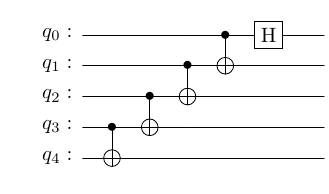}
}%
\subfloat[A $5$-qubit circuit for $\sigma_-^{\otimes 2}\otimes\sigma_+^{\otimes 3}+\sigma_+^{\otimes 2}\otimes\sigma_-^{\otimes 3}$.\label{fig:app:cnot_5q1}]{%
\includegraphics[width=0.5\linewidth]{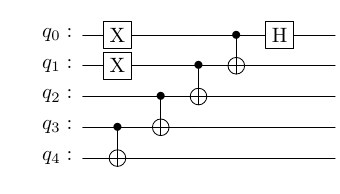}
}%
\caption{Quantum circuit diagonalizing $\sigma_\pm$ polynomials. The $X$-gates are used to convert the circuit (b) to a standardized form (a) ready for the chained $\CNOT$ $\sigma_\pm$ order reduction. $H$-gate diagonalizes the trailing first qubit $X$-gate.}
\label{fig:cnot_diag5q}
\end{figure}

In this section we show some explicit examples of diagonalization of $\sigma_\pm$ polynomials, that naturally appears in the decomposition of the stiffness matrix $\bbK$, \eqref{eq:decomposed_rigidity_K}.
Let us explicitly introduce the matrix notation we are using. We recall the expression of these common matrices:
\begin{align} \label{eq:app:list_matrix}
    p_+=\ket{0}\bra{0}=
    \begin{pmatrix}
        1&0\\
        0&0
    \end{pmatrix};   \quad 
    \sigma_+=\ket{0}\bra{1}=
    \begin{pmatrix}
        0&1\\
        0&0
    \end{pmatrix};  \quad
    \\ \nonumber 
    p_-=\ket{1}\bra{1}=
    \begin{pmatrix}
        0&0\\
        0&1
    \end{pmatrix}; \quad 
    \sigma_-=\ket{1}\bra{0}=
    \begin{pmatrix}
        0&0\\
        1&0
    \end{pmatrix}.  \quad
\end{align}
\begin{equation}
\CNOT=\begin{pmatrix}
    1 & 0 & 0 & 0 \\
    0 & 1 & 0 & 0 \\
    0 & 0 & 0 & 1 \\
    0 & 0 & 1 & 0 \\
\end{pmatrix};
\quad
\sigma_+\otimes\sigma_+=\begin{pmatrix}
    0 & 0 & 0 & 1 \\
    0 & 0 & 0 & 0 \\
    0 & 0 & 0 & 0 \\
    0 & 0 & 0 & 0 \\
\end{pmatrix};
\quad
\sigma_+\otimes p_+=\begin{pmatrix}
    0 & 0 & 1 & 0 \\
    0 & 0 & 0 & 0 \\
    0 & 0 & 0 & 0 \\
    0 & 0 & 0 & 0 \\
\end{pmatrix}.
\end{equation}
The first set of transformations is a partial diagonalization of $\sigma_+\otimes\sigma_+$:
\begin{equation}
\CNOT\cdot\sigma_+\otimes\sigma_+\cdot \CNOT=\begin{pmatrix}
    1 & 0 & 0 & 0 \\
    0 & 1 & 0 & 0 \\
    0 & 0 & 0 & 1 \\
    0 & 0 & 1 & 0 \\
\end{pmatrix}
\cdot
\begin{pmatrix}
    0 & 0 & 0 & 1 \\
    0 & 0 & 0 & 0 \\
    0 & 0 & 0 & 0 \\
    0 & 0 & 0 & 0 \\
\end{pmatrix}
\cdot
\begin{pmatrix}
    1 & 0 & 0 & 0 \\
    0 & 1 & 0 & 0 \\
    0 & 0 & 0 & 1 \\
    0 & 0 & 1 & 0 \\
\end{pmatrix}
=
\begin{pmatrix}
    0 & 0 & 1 & 0 \\
    0 & 0 & 0 & 0 \\
    0 & 0 & 0 & 0 \\
    0 & 0 & 0 & 0 \\
\end{pmatrix}=
\sigma_+\otimes p_+
\end{equation}
Trivially for $\sigma_+^{\otimes3}\equiv\sigma_+\otimes\sigma_+\otimes\sigma_+$ we convert $\sigma_+$ into the diagonal matrix $p_+$:
\begin{equation}
I\otimes\CNOT\cdot\sigma_+^{\otimes3}\cdot I\otimes\CNOT=
\sigma_+\otimes\sigma_+\otimes p_+
\end{equation}
Two consecutive $\CNOT$ transformations lead to two reductions in the number of tensored non-diagonal contributions $\sigma_+$:
\begin{equation}
\CNOT\otimes I \cdot I\otimes\CNOT\cdot\sigma_+^{\otimes3}\cdot I\otimes\CNOT \cdot \CNOT\otimes I =
\sigma_+\otimes p_+\otimes p_+
\end{equation}
Similarly, $n$ consecutive such $\CNOT$ on consecutive qubit couples as shown in Fig. \ref{fig:app:cnot_5q} will transform $\sigma_+\tp n$ into $p_+\tp n$.

For any power of $\sigma_+$, we can define the unitary transformation ${\mathbb{B}}(n)\in \mathcal{M}_{2^n}({\mathbb{R}})$ so that:
\begin{equation}\label{eq:app:sigma_+n}
\mathbb{B}^\dagger(n)
\cdot\sigma_+^{\otimes n}\cdot 
\mathbb{B}(n)=
\sigma_+\otimes p_+^{\otimes (n-1)},\quad \text{where} \quad \mathbb{B}(n)\equiv \prod_{k=0}^{n-2} I^{\otimes(n-2-k)}\otimes\CNOT\otimes I^{\otimes k}.
\end{equation}
In the quantum circuit representation, the ${\mathbb{B}}(n)$ operator corresponds to the consecutive execution of $\CNOT$ gates starting from the last (bottom) qubit, as shown in (Fig. \ref{fig:cnot_diag5q}). The equivalent equation can be obtained for $\sigma_-$ polynomials via the hermitian conjugation of \eqref{eq:app:sigma_+n}:
\begin{equation}
\mathbb{B}^\dagger(n)
\cdot\sigma_-^{\otimes n}\cdot 
\mathbb{B}(n)=
\sigma_-\otimes p_+^{\otimes (n-1)}
\end{equation}

The hermitian operator $\sigma_+^{\otimes n}+\sigma_-^{\otimes n}$ can be presented in the following almost diagonal form, where only the first qubit operator is still off-diagonal:
\begin{equation}
\mathbb{B}^\dagger(n)
\cdot(\sigma_+^{\otimes n} + \sigma_-^{\otimes n})\cdot
\mathbb{B}(n) =
(\sigma_+ + \sigma_-)\otimes p_+^{\otimes (n-1)} \equiv X\otimes p_+^{\otimes (n-1)}.
\end{equation}
We notice that another type of hermitian operator constructed from the antisymmetric matrix $i(\sigma_+^{\otimes n}-\sigma_-^{\otimes n})$ can also be reduced in a similar way, resulting in trailing $Y$ gate:
\begin{equation}
\mathbb{B}^\dagger(n)
\cdot(\sigma_+^{\otimes n} - \sigma_-^{\otimes n})\cdot
\mathbb{B}(n) =
(\sigma_+ - \sigma_-)\otimes p_+^{\otimes (n-1)} \equiv iY\otimes p_+^{\otimes (n-1)}.
\end{equation}
The single qubit gate $X$ and $Y$ can be diagonalized with $H = \usqd \bp 1 & 1\\ 1 & -1\bpp$ and $N = \usqd \bp -1 & -i\\ i & 1\bpp$ respectively, leading to the full diagonalization circuit:
\begin{equation}
\mathbb{B}^\dagger_H(n)
\cdot(\sigma_+^{\otimes n} + \sigma_-^{\otimes n})\cdot
\mathbb{B}_H(n) =
Z\otimes p_+^{\otimes (n-1)};\quad \mathbb{B}^\dagger_N(n)
\cdot i(\sigma_-^{\otimes n} - \sigma_+^{\otimes n})\cdot
\mathbb{B}_N(n) =
Z\otimes p_+^{\otimes (n-1)},
\end{equation}
where we define the unitary of consecutive action of the chained $\CNOT$ and first qubit rotation: $\mathbb{B}_H(n)=\mathbb{B}(n)\cdot H\otimes I^{\otimes (n-1)}$ and $\mathbb{B}_N(n)=\mathbb{B}(n)\cdot N\otimes I^{\otimes (n-1)}$.

The $\CNOT$-chain action on the polynomial containing both $\sigma_+$ and $\sigma_-$ leads also to the shape with a single qubit off-diagonal. We present here only two  hermitian operators that are most relevant for our computations:
\begin{align}
\mathbb{B}^\dagger_H(n+1)\cdot
(\sigma_-\otimes\sigma_+^{\otimes n} + \sigma_+\otimes\sigma_-^{\otimes n})
\cdot \mathbb{B}_H(n+1) =& Z\otimes p_-\otimes p_+^{\otimes (n-1)} \\
i\mathbb{B}^\dagger_N(n+1)\cdot
(\sigma_-\otimes\sigma_+^{\otimes n} - \sigma_+\otimes\sigma_-^{\otimes n})
\cdot \mathbb{B}_N(n+1) =& Z\otimes p_-\otimes p_+^{\otimes (n-1)}
\end{align}
In the general case, any mixed $\sigma_\pm$ powers can be treated with help of supplementary $X$-gate conversion of the initial operator into a standardized form $\sigma_+^{\otimes n}+\sigma_-^{\otimes n}$ prior to diagonalization, illustrated in some examples in Fig.~\ref{fig:cnot_diag5q}.

Having explicitly explained the decomposition of  $\bK{}$ as a sum of matrices in $\mathcal S$, we can now estimate the number of terms that have to be measured separately to calculate $\braket{\psi \mid \bK{} \mid \psi}$ and the diagonalization circuits for each such term.  

Terms of the form $w\otimes I$ and $w \otimes Z$ can be computed simultaneously. However, for terms ending in $X$ (respectively, $Y$), the last qubit has to be once again diagonalized with $H$ (or $N$ respectively). We therefore have to consider such terms separately. The full decomposition and its diagonalization terms are summarized in the table in Sec. \ref{sec:table}.

\newpage
\subsection{Diagonalizable terms from the decomposition of \texorpdfstring{$\bK{}$}{K}}
\label{sec:table}

We display here the $(2n_x n_y +2 n_x + 2 n_y +2)$ terms of a decomposition of $\bbK$, which proves the scaling for the independent measurement of $\bbK$. The tensorial product $\otimes$ has been omitted for clarity. This number of terms can be further reduced by noticing that some of these terms are simultaneously diagonalizable -- for instance, for any $k,k'$, any term diagonalized by $I\tp{n_y+k} \otimes\mathbb{B}_H(n_x-k)\otimes I$ is also diagonalized by $I\tp k \otimes \mathbb{B}_H(n_y-k')\otimes \mathbb{B}_H(n_x-k) \otimes I$.
\def\mmm{\frac{3+4\nu}6}
\def\mmmp{\frac{3+4\nu}{12}}
\def\mmmpp{\frac{3+4\nu}{24}}
\def\mms{\frac{1}8}
\def\idd{\frac{4\nu-1}8i}
\begin{center}
\[
\scalemath{1}{\begin{array}{|c|c|c|}\hline
\text{Term in $\bbK$ decomposition}& \text{Diagonalized by}& \text{\# terms}\\\hline
\mmm\left(4 I \tp {n-1}   - 2p_-\tp {n_y} I\tp {n_x}- 2p_+\tp {n_y} I\tp {n_x} - 2I\tp {n_y} p_-\tp {n_x}- 2I\tp {n_y} p_+\tp {n_x} \right.& \text{\raisebox{-.62em}{$I \tp n$}} &\text{\raisebox{-.62em}{$1$}}  \\
\left.+ p_-\tp {n_y} (p_+\tp{n_x}+p_-\tp{n_x}) +p_+\tp {n_y}(p_+\tp{n_x}+ p_-\tp{n_x})\right)I &&\\\hline
 -\mmmp \sum\limits_{k=0}^{n_y-1}\sum\limits_{k'=0}^{n_x-1} I\tp{k}(\sig_+\sig_-\tp{n_y-k-1}+\sig_-\sig_+\tp{n_y-k-1})  &I\tp k \otimes \mathbb{B}_H(n_y-k)   &\text{\raisebox{-.62em}{$n_xn_y$}}\\
  I\tp{k'}(\sig_+\sig_-\tp{n_x-k'-1}+\sig_-\sig_+\tp{n_x-k'-1})I &\otimes I\tp {k'} \otimes\mathbb{B}_H(n_x-k') \otimes I& \\\hline
\sum\limits_{k=0}^{n_x-1}(p_+\tp{n_y}+p_-\tp{n_y} - 2 I\tp{n_y}) I ^{\otimes k}  (\sigma_+  \sigma_-^{\otimes n_x-k-1} +\sigma_-  \sigma_+^{\otimes n_x-k-1})(\mmmpp I + \mms Z) & I\tp{n_y+k} \otimes\mathbb{B}_H(n_x-k)\otimes I& n_x\\\hline 
\sum\limits_{k=0}^{n_y-1}I\tp k(\sigma_+\sig_-^{\otimes n_y-k-1} + \sigma_-\sig_+^{\otimes n_y-k-1})(p_+\tp{n_x} + p_-\tp{n_x} - 2 I\tp{n_x})(\mmmpp I - \mms Z) & I \tp k \otimes\mathbb{B}_H(n_y-k) \otimes I \tp {n_x+1}& n_y\\\hline
\mms\left(p_+\tp{n_y} - p_-\tp{n_y} \right)\left(p_+\tp{n_x}-p_-\tp{n_x}\right)X &I\tp {n-1}\otimes H & 1\\\hline
\mms\sum\limits_{k=0}^{n_y-1}\sum\limits_{k'=0}^{n_x-1} [ I\tp k \left(\sig_+\sig_-\tp{n_y-k-1}-\sig_-\sig_+\tp{n_y-k-1}  \right)  &  I\tp k \otimes\mathbb{B}_N(n_y-k)   & \text{\raisebox{-.58em}{$n_xn_y$}} \\
 I \tp {k'} (\sig_-\sig_+\tp{n_x-k'-1} - \sig_+\sig_-\tp{n_x-k'-1} )] X &\otimes I\tp {k'} \otimes \mathbb{B}_N(n_x-k')\otimes H &\\\hline
 \idd\sum\limits_{k=0}^{n_x-1} (p_+\tp{n_y} - p_-\tp{n_y})  I ^{\otimes k}\left( \sigma_+  \sigma_-^{\otimes n_x-k-1}  -  \sigma_-  \sigma_+^{\otimes n_x-k-1}\right)Y & I\tp{n_y+k}  \otimes \mathbb{B}_N(n_x-k)\otimes N& n_x\\\hline
 \idd\sum\limits_{k=0}^{n_y-1} I ^{\otimes k}\left( \sigma_-  \sigma_+^{\otimes n_y-k-1}  -  \sigma_+  \sigma_-^{\otimes n_y-k-1}\right) (p_+\tp{n_x} - p_-\tp{n_x}) Y &  I \tp k \otimes \mathbb{B}_N(n_y-k) \otimes I \tp {n_x} \otimes N & n_y\\\hline
\end{array}}\]
\end{center}

\end{document}